\documentclass[a4paper,fleqn,usenatbib]{mnras}
\pdfoutput=1
\usepackage{graphicx}
\usepackage{epstopdf}
\usepackage{color}   
\usepackage[T1]{fontenc}
\usepackage{ae,aecompl}
\usepackage{amssymb,amsmath}
\usepackage{float}
\usepackage{booktabs}
\usepackage{multirow}
\title[Orbital stability of highly eccentric exoplanets]{Orbital stability of coplanar two-planet exosystems with high eccentricities}

\author[K. I. Antoniadou and G. Voyatzis]{
Kyriaki I. Antoniadou,$^{1}$\thanks{E-mail: kyant@auth.gr}
and George Voyatzis$^{1}$
\\
$^{1}$Department of Physics, Aristotle University of Thessaloniki,  Thessaloniki 54124, Greece}

\date{Accepted XXX. Received YYY; in original form ZZZ}

\pubyear{2016}

\begin{document}
\label{firstpage}
\pagerange{\pageref{firstpage}--\pageref{lastpage}}
\maketitle

\begin{abstract}
The long-term stability of the evolution of two-planet systems is considered by using the general three body problem (GTBP). Our study is focused on the stability of systems with adjacent orbits when at least one of them is highly eccentric. In these cases, in order for close encounters, which destabilize the planetary systems, to be avoided, phase protection mechanisms should be considered. Additionally, since the GTBP is a non-integrable system, chaos may also cause the destabilization of the system after a long time interval. By computing dynamical maps, based on Fast Lyapunov Indicator, we reveal regions in phase space with stable orbits even for very high eccentricities ($e>0.5$). Such regions are present in mean motion resonances (MMR). We can determine the position of the exact MMR through the computation of families of periodic orbits in a rotating frame. Elliptic periodic orbits are associated with the presence of apsidal corotation resonances (ACR). When such solutions are stable, they are associated with neighbouring domains of initial conditions that provide long-term stability. We apply our methodology so that the evolution of planetary systems of highly eccentric orbits is assigned to the existence of such stable domains. Particularly, we study the orbital evolution of the extrasolar systems HD 82943, HD 3651, HD 7449, HD 89744 and HD 102272 and discuss the consistency between the orbital elements provided by the observations and the dynamical stability. 
\end{abstract}


\section{Introduction}
Hitherto unprecedented detections of exoplanets have been triggered by missions and ground based telescopes. Analysis of observational data strives to yield orbital elements of planets through various fitting methods. Surprisingly, it is found that    
a large number of exoplanets should have highly elliptical orbits, as recorded in databases \citep[e.g.][]{schn11,han14}. Various mechanisms have been proposed in terms of justifying how planetary orbits became eccentric starting (probably) from almost circular orbits, e.g. planet-planet scattering \citep{Ford05, beaune12} and planetary differential migration \citep{leepeal02,chiang02}. For planets close to their star, although tidal effects result in the circulation of eccentric orbits, when these effects are combined with gravitational interactions with an outer companion, the pumping of the eccentricity is possible \citep{corr12}. Also, tidal-Kozai migration may cause large eccentricity amplitude variation \citep{wumur03,cor11}. 

On the other hand, it has been proposed that high eccentricities may be caused by a wrong interpretation of the observational data and, actually, instead of concluding the existence of a single eccentric planet, it is possible to report that the system hosts two planets in nearly circular motion \citep[e.g.][]{anglada10,wit13,kurster15}. Nevertheless, we cannot argue over the fact that there exist many planets with very eccentric orbits, e.g. HD 80606b, HD 43197b, GJ 317c, HD 45350b, HD 96167b etc.

A single planet system, where tidal effects from the star or from a disc of dust and gas are negligible, does not raise an issue of orbital stability independently of the eccentricity. However, questions for orbital stability arise directly when we assume a possible existence of a second planet in the system, which causes a significant gravitational interaction with the first planet and vice-versa. The three body problem (TBP), which can model efficiently the evolution of a two-planet system, shows significant instabilities when either the planets are very massive or the orbits are so elongated that imply close encounters. Furthermore, the model is not integrable and chaos, either strong or weak, dominates in its phase space. Stability criteria, like the existence of Hill's stability spheres \citep{glad93,giu13,vemu13} are not applied to close-in orbits. Thus, in cases of strong gravitational interactions we get, after a short or a long time interval,  collisions or planet-planet scatterings that eject one of the planets out of the planetary system converting it into an orphan planet.         
Efficient mechanisms that can minimize the effects of mutual gravitational interactions between the two planets and provide long-term stability are the resonances. In many cases, the planetary dynamics shows secular resonances, i.e. the angle $\Delta\varpi$ between the line of apsides is stationary in average, $<\Delta \dot\varpi>=0$  \citep[see e.g.][]{malho02}. More generally, we may refer to apsidal resonances, where $\Delta\varpi$ librates about 0 or $\pi$ (symmetric configurations) or about a different angle (asymmetric configurations) \citep{fmbc05}. The mean motion resonance (MMR) is also an efficient mechanism for orbital stability even for planetary orbits that intersect. MMRs are described by commensurabilities between planetary mean motions, $n_1$ and $n_2$, namely $k_1\,n_1+k_2\,n_2\approx 0$ with $k_i \in\mathbb Z ^*$, and can provide phase protection against close encounters. A large portion of the detected multi-planet systems seems to contain planets that are locked in MMR and a lot of dynamical studies can be found in literature either for particular resonances or particular exosolar systems \citep[see e.g.][]{gozd06,cor09,lask09,coue10,camp11}.

In the present study, we show that apsidal resonances or, mainly, MMR are associated with regions in the phase space of the TBP, where orbits evolve regularly and therefore, will stay in the particular configuration practically forever. These regions can be associated with planetary orbits evolving in high eccentricities or orbits with high amplitude eccentricity oscillations. In literature, eccentricities larger that 0.1 may be characterised as ``high'' or ``moderate''. In our case, we consider orbits with eccentricities larger than 0.3 and we refer to orbits as ``highly eccentric'' when the eccentricity takes values larger than $0.5$. In this context, single planetary systems, like HD 45350b or HD 96167b mentioned above, may host an additional planet without loss of long-term stability.

The aim of this paper is to determine the initial conditions that provide regular evolution for systems with eccentric planets.  In Section 2, we review the model and the basic notions that we use in our analysis. In Section 3, we provide results of an exploration of the phase space, which correspond to eccentric planetary orbits. In Section 4, we analyse the dynamics of exoplanetary systems HD 82943, HD 3651, HD 7449, HD 89744 and HD 102272 which seem to possess at least one planet with high eccentricity. Finally, we discuss the main conclusions of our study. 

\section{Model and Methodology}
The evolution of a two-planet system, whose planets revolve quite far from the star (i.e. the involved bodies are well approximated by point masses), is efficiently modelled by the general three body problem (GTBP). We let two planets, $P_1$ and $P_2$ of masses $m_1$ and $m_2$, revolve around a star, $S$, of mass $m_0$, under their mutual gravitational attraction, with $m_0\gg m_{1,2}$. We will restrict our study to co-planar planetary orbits, which, in the inertial frame, correspond to almost Keplerian ellipses described by heliocentric osculating elements, namely the semimajor axes, $a_i$, the eccentricities, $e_i$ and the longitudes of pericentre, $\varpi_i$. For the position of the planets on the osculating ellipse, we consider the mean anomalies, $M_i$ or the mean longitudes $\lambda_i=M_i+\varpi_i$. Subscripts 1 and 2 shall always refer to the inner and the outer planet, respectively.
  
We performed the numerical integrations of the equations of motion by using the Bulirsch-Stoer integrator. In the equations of motion, the total mass of the system and the gravitational constant are normalized to unity, $\sum_{i=0}^2 m_i =1$. Also, without loss of generality, the initial semimajor axis of the inner planet is always set to $a_1(0)=1.0$. 

A wide exploration of stable trajectories in the phase space of the system requires an enormous number of numerical integrations.  Long-term stability of trajectories is concluded by the computation of {\em chaotic indices}, as set forth below.  The search of stability in our study is also guided by the linearly stable periodic solutions, as described in the following.       

\subsection{Periodic orbits in a rotating frame}\label{model}
For the study of the dynamics of the planar GTBP, we consider a rotating frame of reference $Oxy$ defined in the following way: $O$ is the centre of mass of $S$ and $P_1$, the axis $Ox$ is defined by the direction $S-P_1$ and the axis $Oy$ is perpendicular to $Ox$. So, the planet $P_1$ moves only on the axis $Ox$ and its position is defined by the coordinate $x_1$. The planet $P_2$ moves on the plane $Oxy$ and its position is given by ($x_2,y_2$). The angle, $\vartheta$, which defines the orientation of the rotating frame, is an ignorable variable and therefore, we obtain -further to the energy integral- the integral of angular momentum and the system is reduced to three degrees of freedom \citep{hadj06}. 

Starting planet $P_1$ from an apside (i.e. $\dot{x}_1(0)=0$) at position $x_1(0)=x_{10}$, a periodic solution of period $T$ is defined by the conditions
\begin{equation}\begin{array}{lll}
	x_1(T)=x_{10},& x_2(T)=x_{20}, & y_2(T)=y_{20}, \\
	\dot x_1(T)=\dot x_{10} =0, & \dot{x}_2(T)=\dot{x}_{20}, & \dot{y}_2(T)=\dot{y}_{20}, \nonumber
\end{array}\end{equation}
where the subscript 0 indicates the variable at $t=0$. We determine the initial conditions that satisfy the periodic conditions by using differential corrections and particular continuation process (see \citep{avk11} and references there in).

Due to the GTBP symmetry 
$$
\Sigma :  (x,y,t) \rightarrow (x,-y,-t),
$$
if the initial conditions 
\begin{equation}\label{poic1}
x_{10},\; x_{20},\; y_{20},\; \dot{x}_{10}=0,\; \dot{x}_{20}, \dot{y}_{20}
\end{equation}
correspond to a periodic orbit, then the initial conditions   			
\begin{equation}\label{poic2}
x_{10},\; x_{20},\; -y_{20},\; \dot{x}_{10}=0,\; -\dot{x}_{20}, \dot{y}_{20}
\end{equation}
also correspond to a periodic orbit. If $y_{20}=0$ and $\dot{x}_{20}=0$, then the initial conditions (\ref{poic1}) and (\ref{poic2}) coincide and the periodic orbit is {\em symmetric}. For a symmetric periodic orbit, the two planets can be found simultaneously at their apsides and be in conjunction, namely $\Delta \varpi=\varpi_2-\varpi_1=0$ or $\pi$ and $M_i=0$ or $\pi$ ($i=1,2$). However, in general, the initial conditions (\ref{poic1}) and (\ref{poic2}) correspond to two different periodic orbits (\textit{asymmetric}), the one out of which is the mirror image of the other with respect to symmetry $\Sigma$ \citep{voyhadj05}. If conditions (\ref{poic1}) correspond to the orbital elements $a_i$, $e_i$, $M_i$ and $\Delta \varpi$, then conditions (\ref{poic2}) correspond to $a'_i=a_i$, $e'_i=e_i$, $\Delta \varpi'=2\pi-\Delta \varpi$ and $M'_{i}=2\pi-M_i$. Periodic orbits (\ref{poic1}) or (\ref{poic2}) are not isolated in phase space. They are continued \textit{mono-parametrically} and form \textit{characteristic curves} or \textit{families} of periodic orbits which are called either symmetric or asymmetric according to the type of their periodic orbits \citep{hen97,hadj06}.   
      
Periodic orbits can be studied with respect to their linear stability  \citep{marchal90,hadjbook06}. Let $\mathbf{\xi}=(\xi_1,...,\xi_6)$ indicate a deviation vector for the system's variables, then the variational equations are defined by
\begin{equation}\label{EqVarEq}
\dot{\mathbf{\xi}}=\textbf{J}(t)\mathbf{\xi} 
\end{equation}
with \textbf{J} being the Jacobian of the equations of motion, which is periodic, since it is computed along a particular periodic solution of period $T$. Let $\Delta (t)$ be a fundamental matrix of solutions of (\ref{EqVarEq}) and $\Delta(T)$  the monodromy matrix. Then, the linear stability analysis is based on the location of the pairs of conjugate eigenvalues with respect to the unit circle on the complex plane. Iff all pairs lie on the unit circle (one pair is always equal to unity, due to the energy integral), then the deviations ${\mathbf{\xi}}(t)$ are bounded and the periodic orbit is linearly \textit{stable}. Otherwise, the periodic orbit is linearly \textit{unstable}.

The planar periodic orbits can also be studied with respect to the linear vertical stability \citep{hen,av12,av13}. Vertical stability implies that by adding a small mutual inclination to the coplanar motion, the overall linear stability is not affected. Also, the critical orbits with respect to vertical stability(\textit{vco}) are the bifurcation points which generate spatial periodic orbits.

\subsection{Mean-motion resonances and periodic orbits}\label{mmr}
Families of periodic orbits are classified as \textit{circular} or \textit{elliptic}.  The circular family is symmetric and consists of almost circular planetary orbits and along it the mean motion ratio, $\frac{n_1}{n_2}$, varies. At MMR, where $\frac{n_1}{n_2}\approx \frac{p+q}{p}$, $p,q\in\mathbb Z^*$, the circular family, either continues smoothly with periodic orbits of increasing eccentricities (when $q=1$), or provides bifurcations to elliptic families (when $q\neq 1$) \citep{hadj06}. The generated elliptic families may continue up to planetary orbits with very high eccentricities and along them the ratio $\frac{n_1}{n_2}$ varies slightly for small planetary masses. Thus, elliptic families consist of resonant periodic orbits and each resonant periodic orbit indicates the {\em exact} position of the  MMR in phase space at a particular energy level. 

Periodic orbits can be approximated by the stationary solutions of an appropriate averaged Hamiltonian which depends on the resonant angle variables \citep{beau03}. 
\begin{equation}\begin{array}{c}
\theta_1=p\lambda_1-(p+q)\lambda_2+q\varpi_1 \\
\theta_2=p\lambda_1-(p+q)\lambda_2+q\varpi_2
\end{array}
\nonumber
\end{equation}
The stationary solutions $\dot \theta_i=0$, $i=1,2$, are called also {\em apsidal corotation resonances} (ACR) in \citet{femibe06} and \citet{mbf06}. Thus, in the framework of the averaging approximation, ACR correspond to periodic orbits in the rotating frame. If the periodic orbit (or the ACR) is stable then neighbouring orbits evolve with $\theta_1$ and $\theta_2$ exhibiting libration. The apsidal difference, $\Delta\varpi=(\theta_2-\theta_1)/q$ also librates in this case. When the ACR is defined by a symmetric periodic orbit, librations take place about 0 or $\pi$. By placing the planets at the pericentre or apocentre and letting them be aligned ($\Delta\varpi$=0) or anti-aligned ($\Delta\varpi=\pi$), we get four symmetric configurations. If $q=odd$ the planetary symmetric configurations are defined by the values ($\theta_1,\theta_2$)=$(0,0)$, $(0,\pi)$, $(\pi,0)$ and $(\pi,\pi)$. If $q=even$ the symmetric configurations are distinguished by the same angle arguments, through the pair ($\theta_3,\theta_1$) \citep{av13}, where
\begin{equation}\begin{array}{c}\theta_3=p\lambda_1-(p+q)\lambda_2+\frac{q}{2}(\varpi_1+\varpi_2).\end{array}
\nonumber
\end{equation}

\begin{figure}
\centering
\includegraphics[width=\columnwidth]{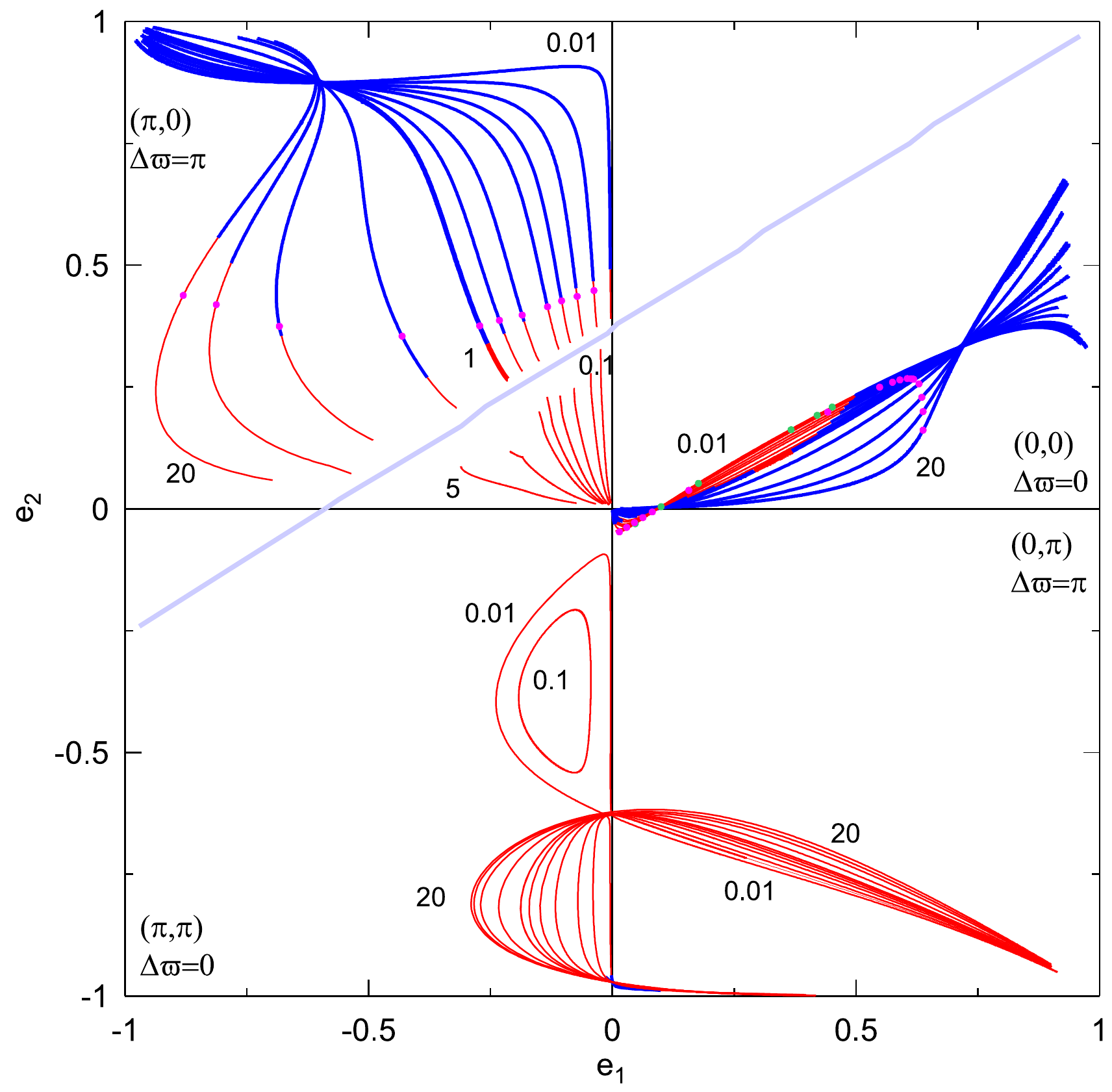}
\caption{Families of symmetric periodic orbits in $2$:$1$ MMR divided in four symmetric configurations, which are distinguished by the resonant angles, $(\theta_1,\theta_2)$, shown in each quadrant. Blue (red) segments represent stable (unstable) periodic orbits. The coloured dots represent the vertical critical periodic orbits. The bold gray curve showcases the region of close encounters between the planets. The numbers in labels correspond to the mass ratio, $\rho$. In computations we used $m_1=0.001$.}
\label{21}
\end{figure}

Along an elliptic symmetric family of periodic orbits the configuration does not change and the family can be represented by characteristic curves on the plane of eccentricities ($e_1$,$e_2$). Positive or negative values are given respectively to the eccentricity of the inner or outer planet, when the resonant angles librate about $0$ or  $\pi$. For small planetary masses, these characteristic curves depend on the mass ratio, $\rho=m_2/m_1$, rather than the individual planetary masses \citep[see e.g.][]{beau03,femibe06}. Also, as we mentioned above, the ratio $\frac{n_1}{n_2}$ varies slightly along the family and therefore the ratio of semimajor axes $\alpha$ varies correspondingly. In Fig. \ref{21}, we present the families of symmetric periodic orbits in the $2$:$1$ MMR for various mass ratio values and we classify them in the four symmetric configurations, where they belong to. 

In the case of asymmetric periodic orbits (i.e. asymmetric ACR), besides the variation of the eccentricities along the family, the resonant angles, $\theta_i$, and $\Delta\varpi$ also vary and generically their values are  different from 0 or $\pi$. Thus, the depiction of the families on the plane ($e_1$,$e_2$) does not provide the values of the resonant angles. 

\subsection{Tracing regions of stability} \label{DSMAPS}
Eccentric planetary orbits avoid possible close approaches \textit{only} when they are located in a particular resonant configuration. However, this is not a sufficient condition for long-term stability, because multitude of weak encounters may cause chaotic evolution, diffusion in phase space and, finally, the disruption of the system, which entails collisions or escape of the planet. On the other hand, \citet{long} have reported chaotic orbits of mutually inclined planets that survive without collisions or escapes for billion years, but these cases should be exceptions.  

\begin{figure}
\centering
\includegraphics[width=\columnwidth]{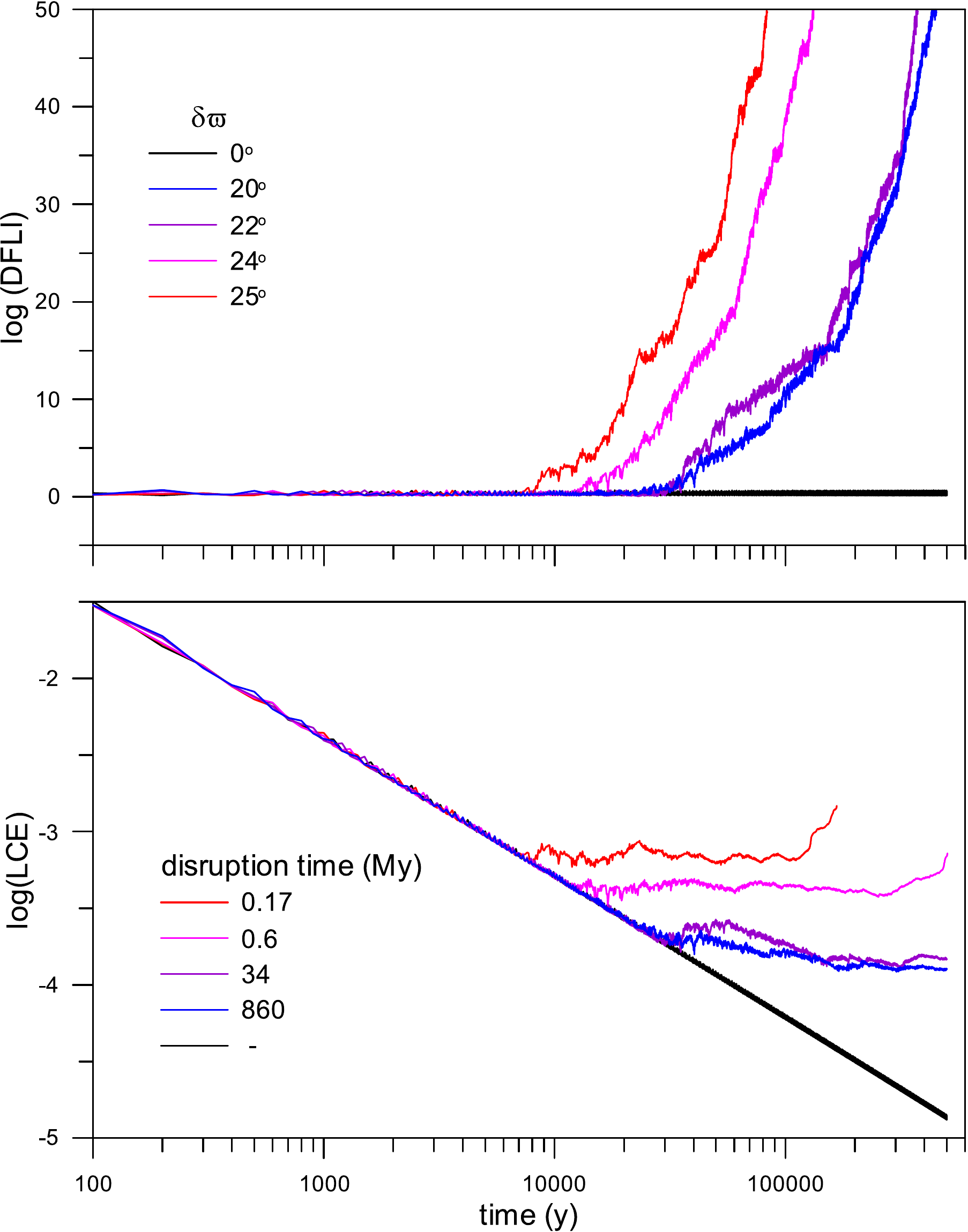}
\caption{The evolution of DFLI and LCE for the system HD 82943 (black curve) with initial conditions given in Table \ref{tab}. The rest cases correspond to the same initial conditions, but the longitudes of pericentre $\varpi_1$ differ from the original value by $\delta\varpi$ given in the top legend. The bottom legend indicates the time when disruption of the system occurs according to our numerical integrations.}
\label{dflilcn}
\end{figure}

In Hamiltonian systems of three degrees of freedom, like the planar GTBP, orbits are either regular (quasi-periodic), winding invariant tori, or chaotic. In regions of phase space occupied mainly by invariant tori, long-term stability is guaranteed, since a possible Arnold diffusion should be too slow. In numerical studies, phase space regions of chaotic or regular evolution can be revealed by computing chaotic indices \citep{maff11}. The most known and well established index is the Lyapunov Characteristic Exponent, LCE \citep{benettin}. Also,  \citet{froe97} introduced the Fast Lyapunov Indicator (FLI) which, may be found in literature in various, but equivalent, definitions. For instance, in \citep{voyatzis08} the {\em de-trended} FLI is used for the study of planetary orbits and is defined as
\begin{equation}                
DFLI(t)=\frac{1}{t} \frac{\xi(t)}{\xi(0)},
\end{equation}
where $\xi$ is the deviation vector of the orbit, which is being evaluated along the numerical integration in tandem with the integration of the linearized system. For a regular orbit DFLI remains almost constant over time, but when chaos is detected, DFLI increases exponentially taking very large values.  In Fig. \ref{dflilcn} (black curves), we present the DFLI and the LCE along the orbit of the system HD 82943b,c by using the initial conditions given in Table \ref{tab}. For the LCE we obtain the usual behaviour shown for regular orbits, namely LCE tends to zero. On the other hand, the DFLI evolves constantly about the value 10. By increasing the angle of the line of apsides of the original system by a $\delta\varpi$, we obtain chaotic orbits which are indicated clearly by the evolution of both the LCE and the DFLI. Particularly, after a critical integration time $t_c$, LCE tends to a non-zero value, whereas DFLI increases rapidly to very large values. In the legend of the figure, we indicate also the time when disruption of the planetary system occurs.  

The computational advantage provided by the DFLI is that we can recognize certainly its large value, stop the numerical integration and classify the orbit as chaotic. In our study, we stop the integration when $DFLI(t)>10^{30}$ or at $t_{max}=200 Ky$. By selecting appropriate plane grids of initial conditions we construct {\em Dynamical Stability maps} (DS-maps) that visualise the value of DFLI at each grid point. White points indicate failure of numerical integration for $t<t_{max}$, due to very small integration step caused by very close encounters.  Generally, the regions which are classified as stable (regular orbits) are sharply distinguished by chaotic regions, which correspond mainly at the largest DFLI value. Numerical tests showed that after increasing the integration time at $t_{max}=1 My$, the dynamical maps remain qualitatively almost invariant. Thus, the value $t_{max}=200 Ky$ is considered efficient in our study for distinguishing chaos from order.                   

For highly eccentric orbits, where close encounters and collisions are probable, chaos dominates. However, appropriate resonant configurations may provide stable orbits. As we mentioned above, such resonant configurations are centred at linearly stable periodic orbits and this is a \textit{guide} for tracing stability regions for high eccentricities.   

\section{Highly eccentric regular orbits}\label{hecc}

In this section, we explore the phase space, in order to get a global view of the regions of regular orbits of high eccentricities and provide a general description. Firstly, we determine the areas of collisional orbits and then, we compute DS-maps. 

\begin{figure}
\centering
\includegraphics[width=\columnwidth]{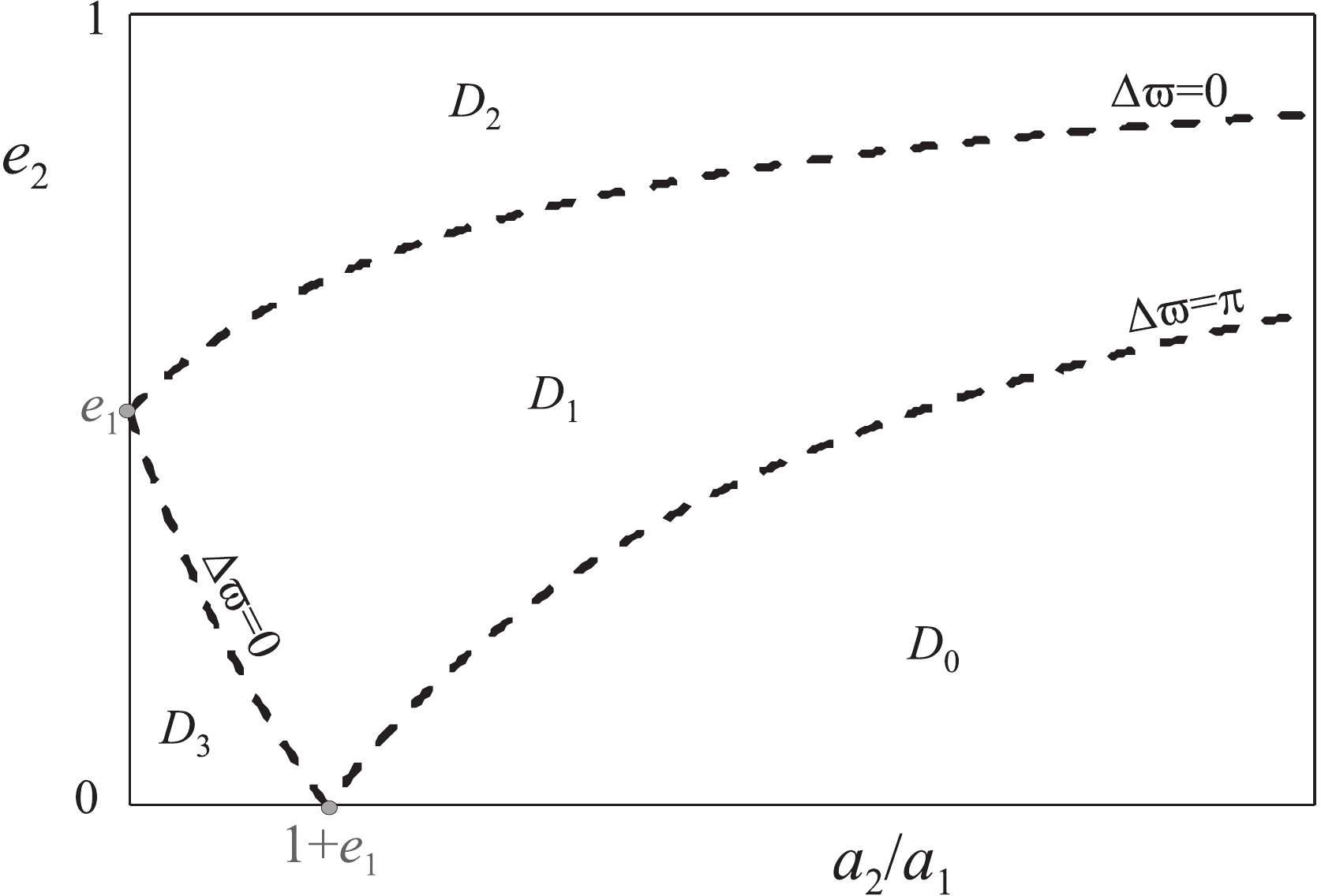}
\caption{Domains, $D_i$, of collisional orbits defined by the inequality in Eq. \eqref{ColCri}. In $D_1$, the appearance of orbital intersections depends on the apsidal angle. In $D_2$ and $D_3$, orbits intersect regardless of their apsidal angle.}
\label{ColPlane}
\end{figure}

Two Keplerian planetary orbits intersect, if the following criterion holds \citep{kho99}
\begin{equation}
a_1^2(1-e_1^2)+a_2^2(1-e_2^2)-2a_1 a_2 (1-e_1 e_2 \cos\Delta \varpi) \leq 0.
\label{ColCri}
\end{equation}
The above criterion actually depends on the ratio of semimajor axes $\alpha$. By using the inequality in Eq. \eqref{ColCri}, assuming for the inner planet a fixed eccentricity value $e_1$ and setting $\Delta\varpi =0$ or $\pi$, we obtain the curves of Fig. \ref{ColPlane} that separate the plane $(\alpha,e_2)$ in four domains, $D_i$, $i=0,..,3$. In $D_0$, the planetary orbits do not intersect for any apsidal angle, $\Delta\varpi$. In $D_1$, anti-aligned orbits intersect and also, orbits with apsidal angle $\Delta\varpi$ around the value $\pi$ that satisfies the inequality Eq. \eqref{ColCri}. In $D_2$ and $D_3$, planetary orbits intersect regardless of their apsidal angle.         

We compute DS-maps, as we have described in Section \ref{DSMAPS}, for a fictitious Jupiter - Saturn system ($m_1=0.001$, $m_2=0.0003$). We consider a grid of initial conditions of size $300\times 100$ on the plane $(\alpha,e_2)$ by considering for the inner planet initial eccentricity values $e_1=0.3$ and $0.6$. We always start the planets from their periastron and we consider the cases of initially aligned or anti-aligned orbits. We remind that in our computations we set $a_1=1$; hence, $\alpha\equiv a_2$.  

\begin{figure*}
\begin{center}
$\begin{array}{cc}
(a)\;\;e_1=0.3,\; \Delta\varpi=0 & (b)\;\;e_1=0.3,\; \Delta\varpi=\pi \vspace{-0.1cm} \\
\includegraphics[width=\columnwidth]{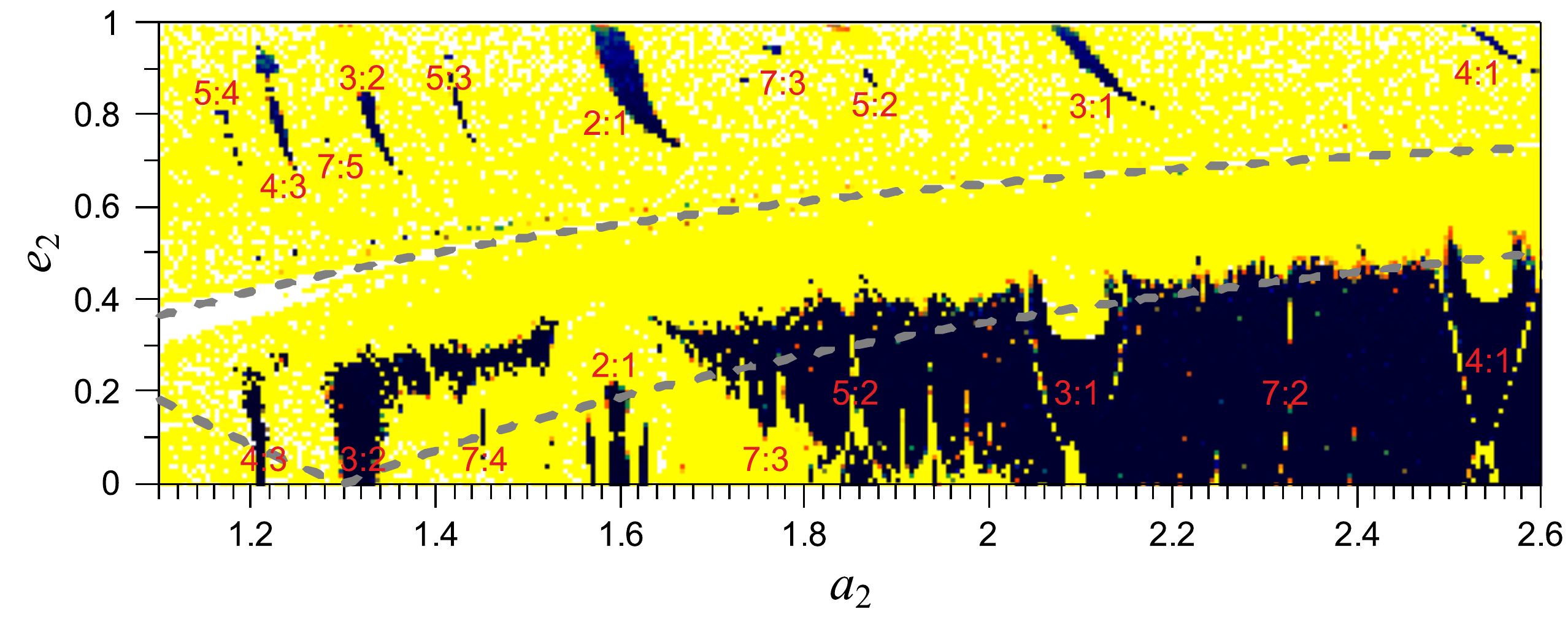} & \includegraphics[width=\columnwidth]{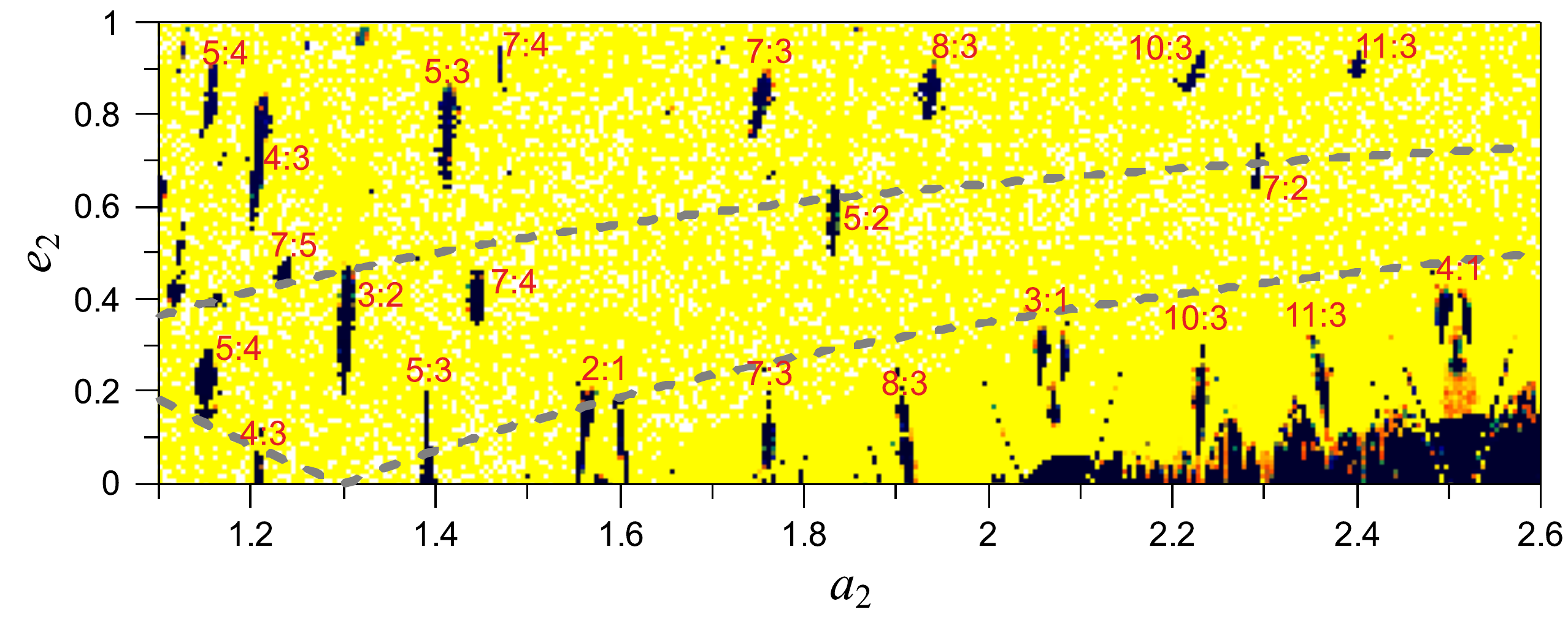}\\
(c)\;\;e_1=0.6,\; \Delta\varpi=0 & (d)\;\;e_1=0.6,\; \Delta\varpi=\pi \vspace{-0.1cm} \\
\includegraphics[width=\columnwidth]{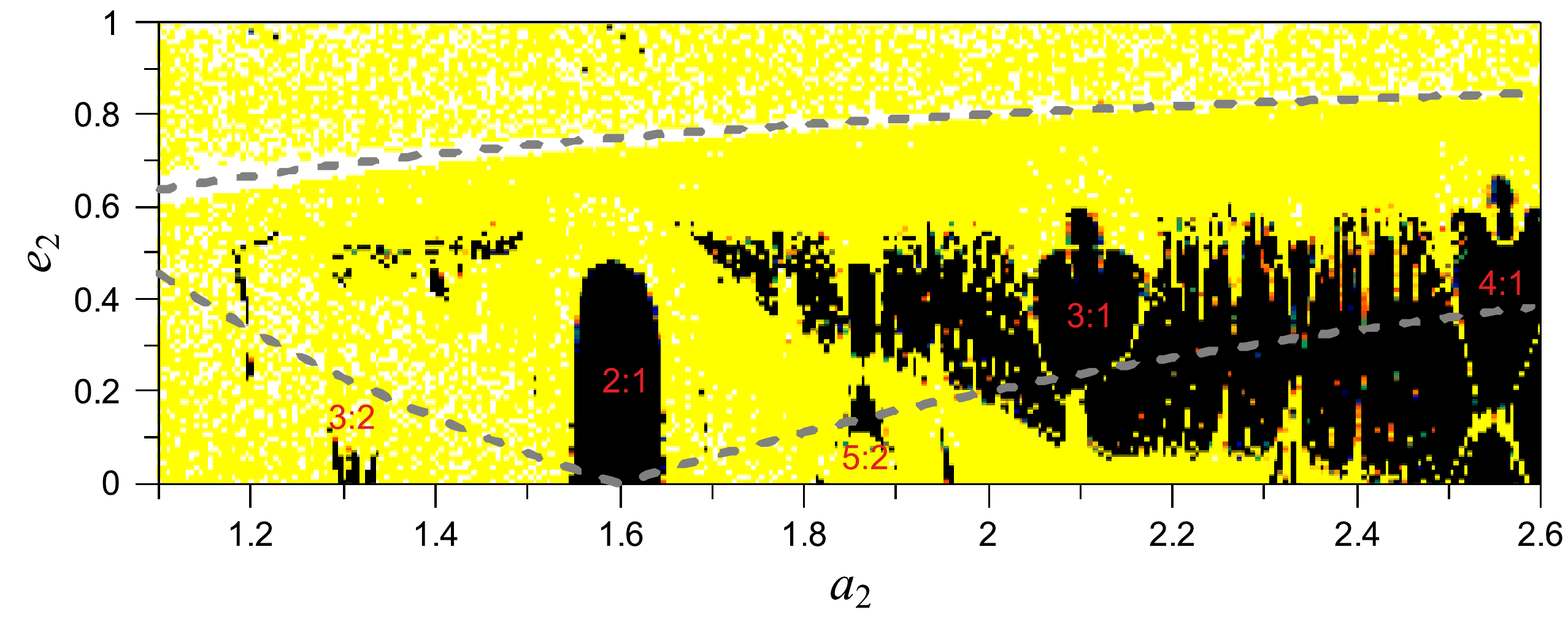} & \includegraphics[width=\columnwidth]{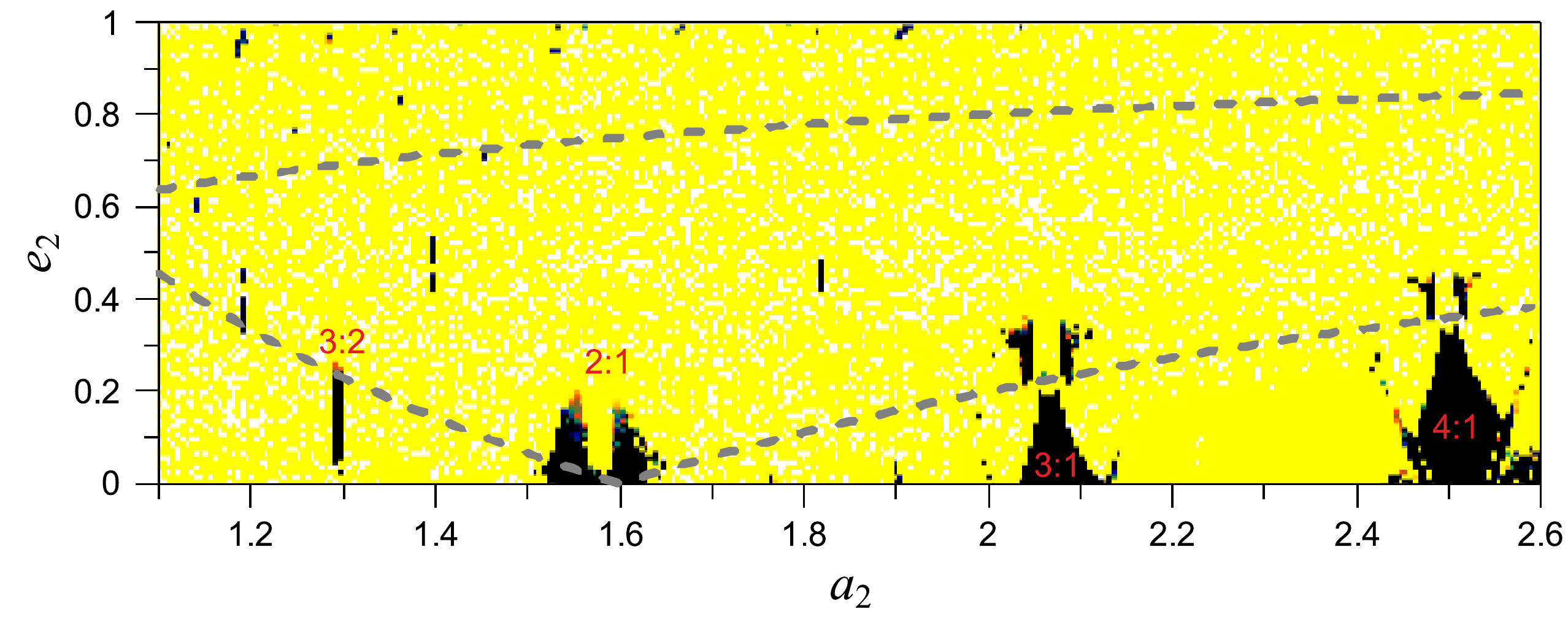}\vspace{-0.3cm}\\
\end{array}$
$\begin{array}{c}
\includegraphics[width=3.cm,height=0.6cm]{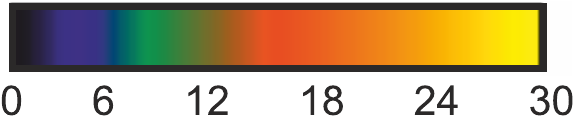} 
\end{array}$
\caption{DS-maps on the plane $(a_2,e_2)$ for $m_1=0.001$, $m_2=0.0003$, snd initial conditions  $e_1=0.3$ (top) and $0.6$ (bottom), $M_1=M_2=0$ and $\Delta\varpi=0$ (left) and $\pi$ (right). Regions of MMR are indicated. Dashed lines indicate the regions of collisions presented in Fig. \ref{ColPlane}. The coloured bar corresponds to the logarithmic values of the DFLI.}
\label{DSMa2e2}
\end{center}
\end{figure*}

\begin{figure}
\centering
\includegraphics[width=\columnwidth]{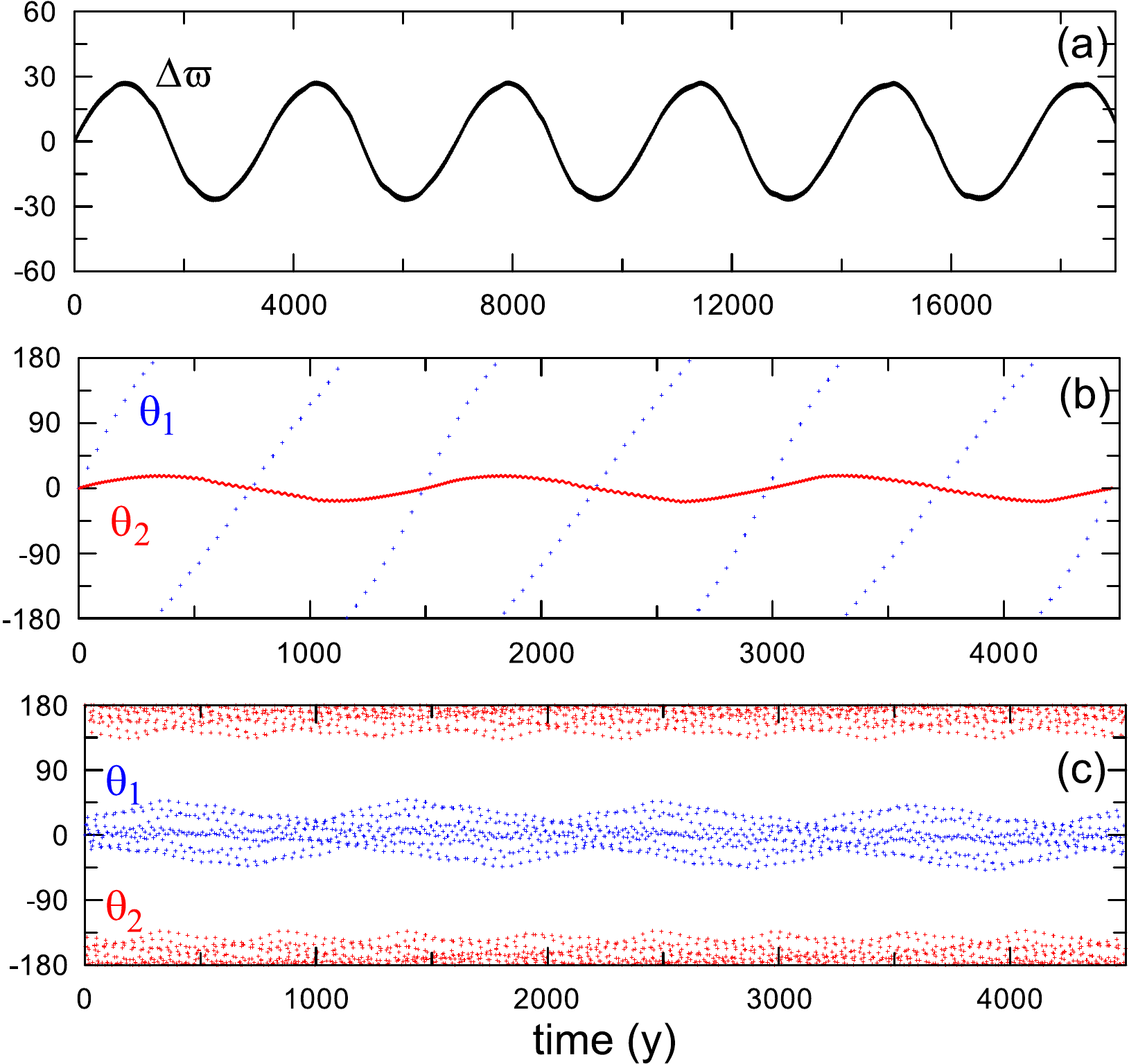}
\caption{\textbf{a} Apsidal resonance: libration of $\Delta\varpi$ for an orbit with initial conditions  $a_2=1.82$, $e_2=0.36$ of the map of Fig. \ref{DSMa2e2}a.  \textbf{b} $2/1$ secondary resonance (between $\theta_1$-rotation and $\theta_2$-libration) in the $3$:$1$ MMR.  \textbf{c} Librations of angles $\theta_1$ and $\theta_2$ around 0 and $\pi$, respectively, near the $8$:$3$ ACR.}
\label{RangEvol}
\end{figure}

For initially aligned orbits the map of Fig. \ref{DSMa2e2}a reveals that in the domain $D_0$ and for about $a_2>1.7$ the majority of orbits, resonant or non resonant, is regular. This regular region seems to extend beyond region $D_1$, which is above the collision line for $\Delta\varpi=\pi$. Although the planetary orbits start as aligned, if $\Delta\varpi$ rotates, then, after a long evolution, they will become anti-aligned and close encounters may take place. However, in these cases, apsidal resonance is present, i.e. $\Delta\varpi$ librates around zero and planetary orbits are protected from a close encounter \citep{malho02}. Actually, all orbits in the domain $D_1$, when not in a MMR, should exhibit apsidal resonance (see Fig. \ref{RangEvol}a for an example). In the domain $D_2$ of very high eccentricities, regular orbits are located in distinct regions, which form islands in the broad chaotic region. All orbits in such islands are in a MMR. Studying the evolution of resonant angles, $\theta_1$ and $\theta_2$, we obtain that in $5$:$4$, $7$:$5$ and $7$:$3$ resonances all angles librate and, consequently, $\Delta\varpi$ librates, too. This indicates that the orbits in these islands should be located near an ACR (i.e. a periodic orbit). In the rest cases, only $\theta_2$ librates, while $\theta_1$ and $\Delta\varpi$ rotate. Considering the average frequencies $f_{1R}$ and $f_{2L}$ of $\theta_1$-rotation and $\theta_2$-libration, respectively,  we observe the existence of resonances $f_{1R}$:$f_{2L}=s_1/s_2$, with $s_i$ being integers, known as \textit{secondary} resonances in asteroid dynamics \citep[e.g.][]{momo93}. 
In planetary dynamics, secondary resonance are reported by \cite{beauge08} inside the $2$:$1$ MMR, but in low eccentricities.
Obviously, if $\theta_2$ librates, $\Delta\varpi$ oughts to rotate with frequency $f_{1R}$. In the islands of $4$:$3$, $3$:$2$ and $2$:$1$ MMR we find the secondary resonance $1/1$. In the islands of $5$:$2$, $3$:$1$ and $4$:$1$ MMR we find the secondary resonances $3/1$, $2/1$ and $3/1$, respectively. An example of a resonant evolution of $\theta_1$-rotation and $\theta_2$-libration is shown in Fig. \ref{RangEvol}b. We should mention that near exact resonances, where resonant angles librate, the eccentricities show variations of small amplitudes. In the rest cases, the eccentricity oscillations are of large amplitude and anti-phase, due to the conservation of the angular momentum deficit \citep{Laskar00}.
    
By assuming that the planetary orbits are initially anti-aligned, we get the map of Fig. \ref{DSMa2e2}b. In this case, regular orbits of high eccentricities are mainly restricted to islands of mean motion resonances. Now, in most cases both $\theta_1$ and $\theta_2$ librate either around $0$ or $\pi$, while the apsidal angle also librates around $\pi$ keeping the orbits almost anti-aligned. An example of such librations in the $8$:$3$ MMR is presented in  Fig. \ref{RangEvol}c. The existence of exact resonant configurations (related to periodic orbits) have been shown for the MMR $4$:$3$, $3$:$2$, $2$:$1$, $5$:$2$, $3$:$1$, $4$:$1$ and $5$:$1$ in \citep{mbf06,av12,av13}.  
              
By setting the initial eccentricity of the inner plane at the high value $e_1=0.6$ we get the DS-maps of Fig. \ref{DSMa2e2}c,d. For initial $\Delta\varpi=0$, most regular orbits are confined to MMR, which are located between thin chaotic zones (\textit{separatrices}). Apsidal resonances are also apparent outside the MMR. We should remark the relatively large region of $2$:$1$ resonant orbits, where resonant angles librate around 0. In the well delineated regions of $3$:$1$ and $4$:$1$ MMR, the corresponding resonant angles rotate and thus, such regions are not associated with the existence of periodic orbits. Despite the case of $e_1=0.3$, islands of regular motion in the domain $D_2$ are almost absent for $e_1=0.6$. By computing the map of initially anti-aligned orbits (panel d), we distinguish regions of regular orbits. The most significant ones are located in the MMRs $2$:$1$, $3$:$1$ and $4$:$1$. The collision line for $\Delta\varpi=\pi$ seems to delimit the above regions, because of the planetary collisions. However, above this line, we obtain the formation of two islands of regular motion (left and right of the centre of the MMR). In these islands,  the resonant angles librate. Additional computations of DS-maps show, that, when we start with anti-aligned orbits and the outer planet is initially located at apastron, then, large islands with librating resonant angles appear in the domain $D_2$.            

The above analysis indicates the existence of many regions of regular orbits, where at least one of the planets starts with very high eccentricity. From a dynamical point of view, the most interesting regular orbits for exoplanets are those which are associated with the existence of families of periodic orbits. Many studies (\citep[e.g.][]{durkley15} and references there in) have showed that the capture of planetary systems in MMRs can occur after differential migration. Periodic orbits provide a \textit{path} for such a migration along the MMR which starts from low eccentricities and may extend up to very large eccentricities \citep{bmfm06,vat14}. In Section \ref{exohecc}, we focus on particular exosystems and attempt to locate them at regular regions near periodic orbits.            

\section{Extrasolar Planetary Systems}\label{exohecc}

We study the long-term dynamical stability of the resonant extrasolar planetary systems HD 82943, HD 3651, HD 7449, HD 89744 and HD 102272, which seem that can possess at least one planet evolving on highly eccentric orbit. We will take into account the data given in Table \ref{tab}. Our aim is to show the efficiency of the phase-protection mechanism provided by MMR and complement, constrain or, even, propose the possible range of the values of the orbital elements, in order for these systems to evolve regularly. Based on this, we break down our method in the following steps:
\begin{enumerate}
  \item  For each particular planetary system, of mass ratio $\rho$ and its possible MMR, we compute the families of symmetric periodic orbits. We divide them into 4 different symmetric configurations and  we use the eccentricities plane,($e_1,e_2$), in order to project their characteristic curves. We also detect and compute families of asymmetric periodic orbits with eccentricities and semimajor axes that may be dynamically relevant to the observations.   
	\item We construct DS-maps in the eccentricity or other planes, by using initial conditions relevant to the observations or periodic orbits (or a combination of them). For the study of symmetric configuration the provided DS-map includes the different configurations in 4 quadrants and the computed symmetric families of periodic orbits are overplotted.     
\end{enumerate}

We mention that we take always as initial conditions $a_1(0)=1$, $\varpi_1(0)=0$ and $M_1=0$, except in cases where these values are defined otherwise.    
 
\begin{table}
\caption{Data for planets taken from {\em{exoplanet.eu}}. The unit of length is normalized such that $a_1=1$. 
}\resizebox{\columnwidth}{!}{%
\begin{tabular}{lccccc}\toprule%@{\hspace{-.15em}}
 &$\rho$ & $a$ (AU) &   $e$  &  $\varpi$ (deg) & $M$ (deg)\\
\midrule
HD 82943 c &\multirow{2}{*}{1.004} &1.000 & 0.425 &   133&  256 \\ 
HD 82943 b & &1.595 & 0.203 &   107&  333 \\ 
HD 3651 b  &\multirow{2}{*}{2.219} &1.000 & $0.04\pm 0.2$ &   $55\pm 66$&  $-$ \\ 
HD 3651 c  & &1.586 & $0.596\pm 0.036$ &   $242.5\pm 4.5$&  $-$ \\ 
HD 7449 b  &\multirow{2}{*}{1.802} &1.000 & $0.82\pm 0.06$ &   $339\pm 6$&  $0$ \\ 
HD 7449 c  & &2.156 & $0.53\pm 0.08$ &   $11\pm 8$&  $307.95\pm13.35$ \\
HD 89744 c &\multirow{2}{*}{2.250} &1.000 & $0.0\pm 0.05$ &   $0\pm 86$&  $-$ \\ 
HD 89744 b & &2.000 & $0.7\pm 0.02$ &   $195\pm 3$&  $-$ \\ 
HD 102272 b &\multirow{2}{*}{0.441} &1.000 & $0.05\pm 0.04$ &   $118\pm 58$&  $12\pm60$ \\ 
HD 102272 c & &2.557 & $0.68\pm 0.06$ &   $320\pm 10$&  $324\pm40$ \\
\bottomrule\end{tabular}}
\label{tab}
\end{table}

\subsection{HD 82943}

The suggested data for HD 82943 are shown in Table \ref{tab} \citep[see also][]{tan13}. The period ratio is $\frac{T_2}{T_1}\approx 2.017$. The normalized planetary masses are $m_1=0.004018$ and $m_2=0.004035$.

\begin{figure}
\begin{center}
\includegraphics[width=\columnwidth]{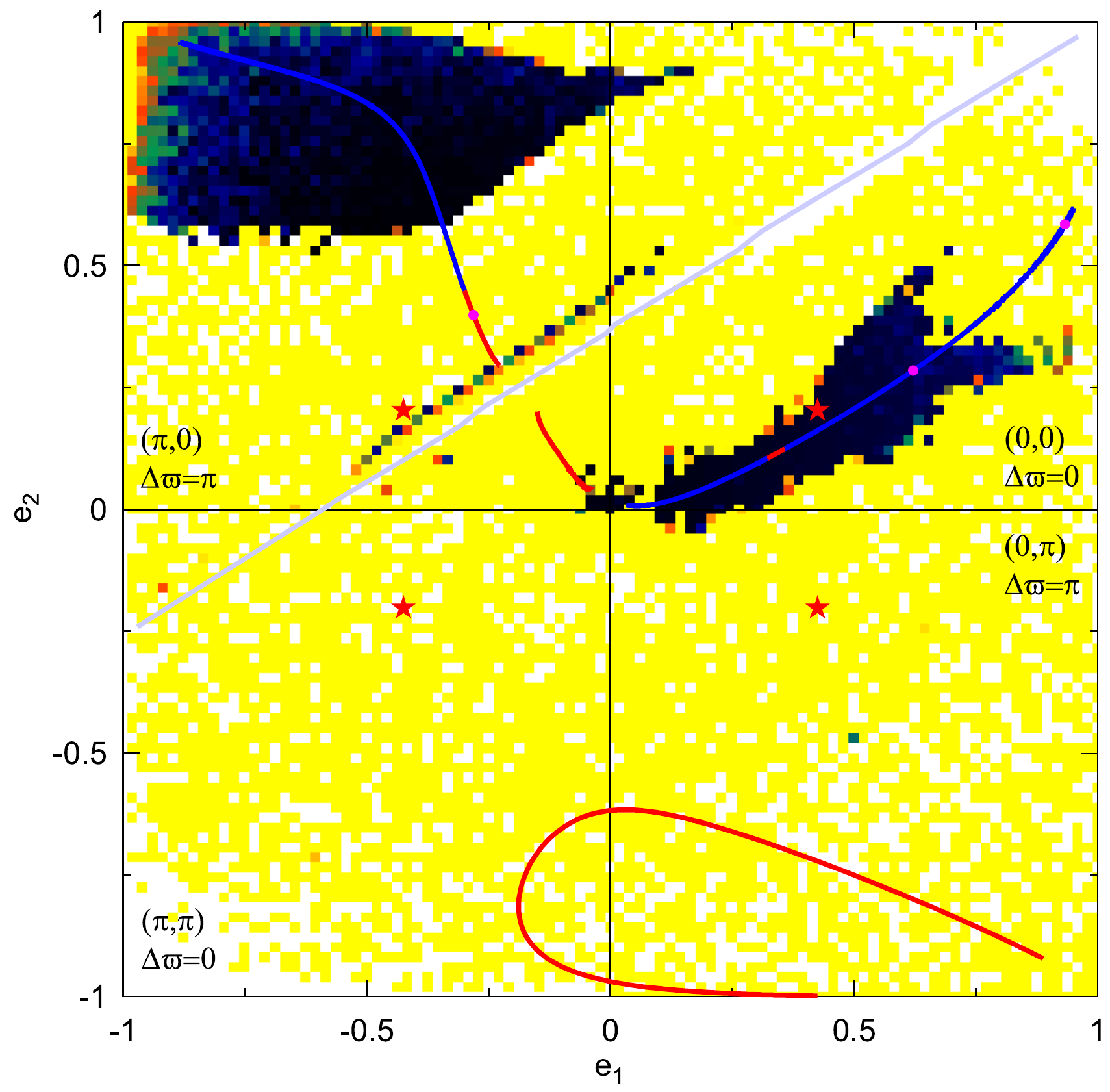}
\caption{DS-map on ($e_1,e_2$) plane of symmetric configurations  computed for the observational value $\alpha=1.595$ of HD 82943. Dark (pale) coloured regions stand for regular (chaotic) evolution. The families of symmetric $2$:$1$ resonant periodic orbits are also presented, as in Fig. \ref{21}. The red stars depict the observational values of HD 82943 given in Table \ref{tab}.}
\label{82943map}
\end{center}
\end{figure}

In Fig. \ref{82943map}, we present the main families of symmetric periodic orbits computed for the system's planetary mass ratio, $\rho=1.004$, which are projected on the $(e_1,e_2)$ plane and divided in the four symmetric configurations. In tandem with these, we fit in the respective DS-maps for each quadrant, which are computed for the observational semimajor axis ratio $\alpha=1.595$. 

\begin{figure}
\begin{center}
\includegraphics[width=\columnwidth]{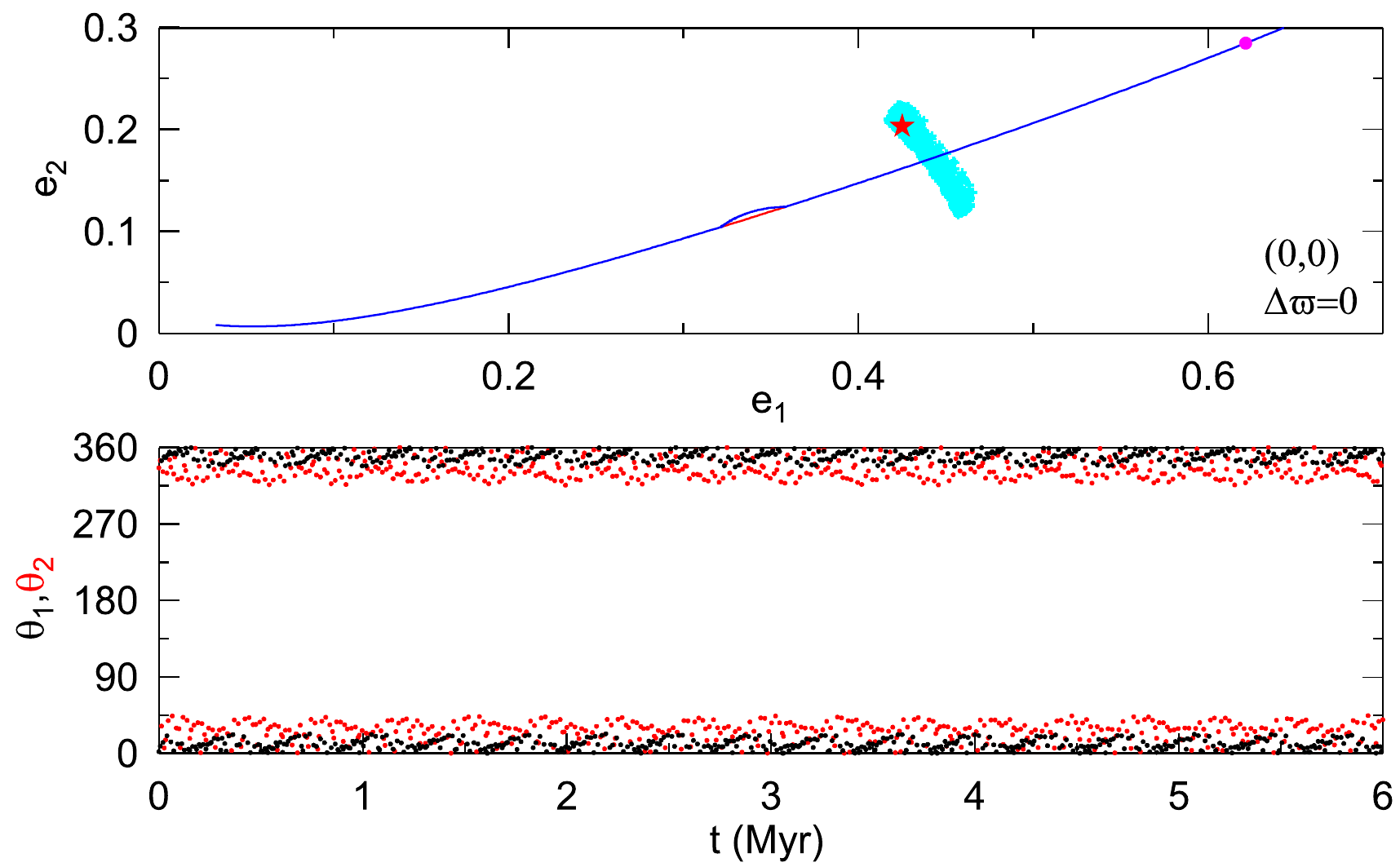}
\caption{The evolution of HD 82943 on the plane of eccentricities (cyan line, top panel), which takes place about a stable symmetric periodic orbit of the configuration $(\theta_1,\theta_2)=(0,0)$. The libration of resonant angles about $0^\circ$ is shown in the bottom panel.} 
\label{82943evol}
\end{center}
\end{figure}

Obviously, HD 82943 is located within the region of stability which belongs to configuration $(0,0)$, since the rest symmetric configurations exhibit chaotic evolution. Starting with the initial conditions given in Table \ref{tab} the resonant angles and the apsidal difference librate about $0^{\circ}$ and the evolution of the system is regular and centred at a stable symmetric periodic orbit $(e_1,e_2)=(0.441,0.171)$ where $\alpha=1.62570$ (see Fig. \ref{82943evol}). Additionally, we should note that the periodic orbits, up to the vertical critical orbit depicted both in Figs. \ref{82943map} and \ref{82943evol}, are vertically stable; hence, we can infer that these planets could also evolve regularly, when small mutual planetary inclination is introduced.

\begin{figure}
\begin{center}
\includegraphics[width=\columnwidth]{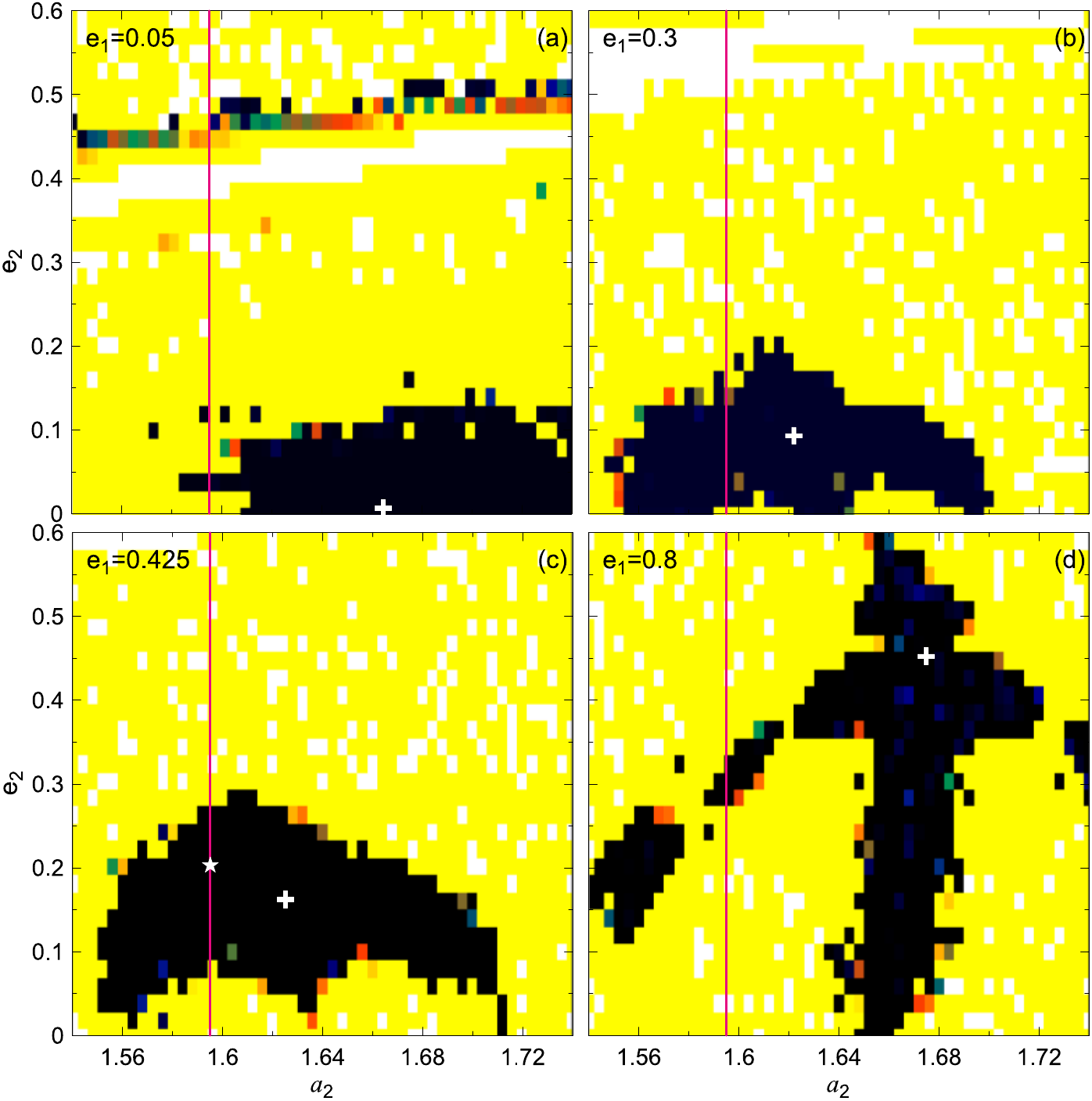}
\caption{DS-maps on the plane $(a_2,e_2)$ for the configuration $(\theta_1,\theta_2)=(0,0)$, and \textbf{a} $e_1=0.05$, \textbf{b} $e_1=0.3$, \textbf{c} $e_1=0.425$, \textbf{d} $e_1=0.8$. The white crosses signify the resonant periodic orbits and the white star the location of HD 82943. The magenta line indicates the $\alpha=a_2/a_1$ value for which the DS-map of Fig. \ref{82943map} was computed.}
\label{82943a2e2}
\end{center}
\end{figure}

Note-worthy is the presence of a family of stable periodic orbits in the configuration $(\pi,0)$, which creates a broad region, where the orbits are regular above the collision line. Finally, we should remark again on the fact that the ratio of the semimajor axes changes slightly along the families. This, in turn, creates the misleading (yellow) region of irregular evolution in the vicinity of stable (blue) periodic orbits, since the observational semimajor axis ratio, for which the DS-map in Fig. \ref{82943map} is computed, was assumed as constant. In Fig. \ref{82943a2e2}, we present the islands of stability, which become apparent on planes $(a_2,e_2)$ around the stable periodic orbits (ACR) of the family in the ($0,0$) configuration.  The dynamical evolution of HD 82943 is also studied in \citep{balbe13} and its stability and resonant properties are indicated, too.  

\subsection{HD 3651}

The suggested data for HD 3651 are shown in Table \ref{tab} \citep[see also][]{wit09,wit13}. The period ratio is $\frac{T_2}{T_1}\approx 2.007$. The masses are normalized as $m_1=0.000114$ and $m_2=0.000253$. \citet{wit13} decrease the eccentricities, $e_i<0.06$, when they perform two-planet fitting, in order for the system to be stable. We hereby show that this system could also be stably evolving about a highly eccentric either symmetric or asymmetric periodic orbit.   

\begin{figure}
\centering
\includegraphics[width=\columnwidth]{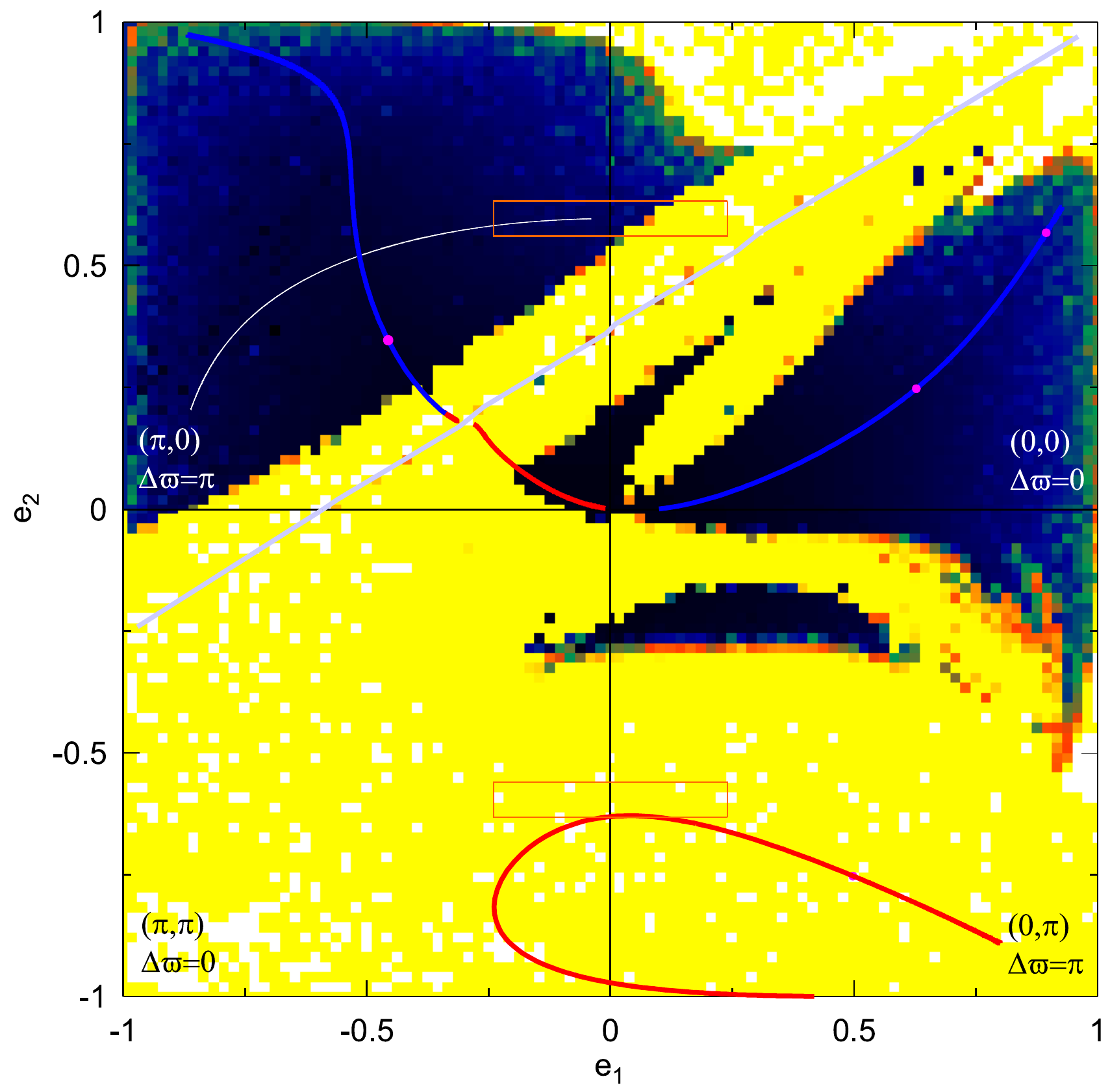}
\caption{DS-map on ($e_1,e_2$) plane of symmetric configurations computed for $\alpha=2^{2/3}$ for the system HD 3651.  Families of symmetric periodic orbits are also presented. Orange boxes represent the possible range of the observational eccentricities.}
\label{3651all}
\end{figure}

In the DS-map computed for the unperturbed approximation of the MMR  ($\alpha=2^{2/3}\approx 1.5874$) that is shown in Fig. \ref{3651all}, we observe that the system HD 3651 could be evolving in ($\pi,0$) symmetric configuration, since the possible eccentricities according to Table \ref{tab} and presented by the orange box in the DS-map, are located in the region of stability, which exists thereby. The boundaries of the stable regions do not change significantly if the observational value $\alpha=1.5860$ is used. By fixing the apsidal difference equal to $187.5^\circ$ (observational value), $M_1=180^\circ$ and $M_2=0^\circ$ (the respective symmetric configuration), the evolution of the system, which is shown by the white curve at the top-left quadrant of Fig. \ref{3651all}, is centred at the stable periodic orbit $(e_1,e_2)=(-0.515,0.522)$ where $\alpha=1.58677$. The oscillations of the eccentricity values are very large, because the system is not located close to this periodic orbit (or the ACR, equivalently). The prohibited symmetric configurations are $(0,0)$, $(0,\pi)$ and $(\pi,\pi)$, because for the system's possible location, the chaotic indicator suggests irregular evolution.

\begin{figure}
\centering
\includegraphics[width=\columnwidth]{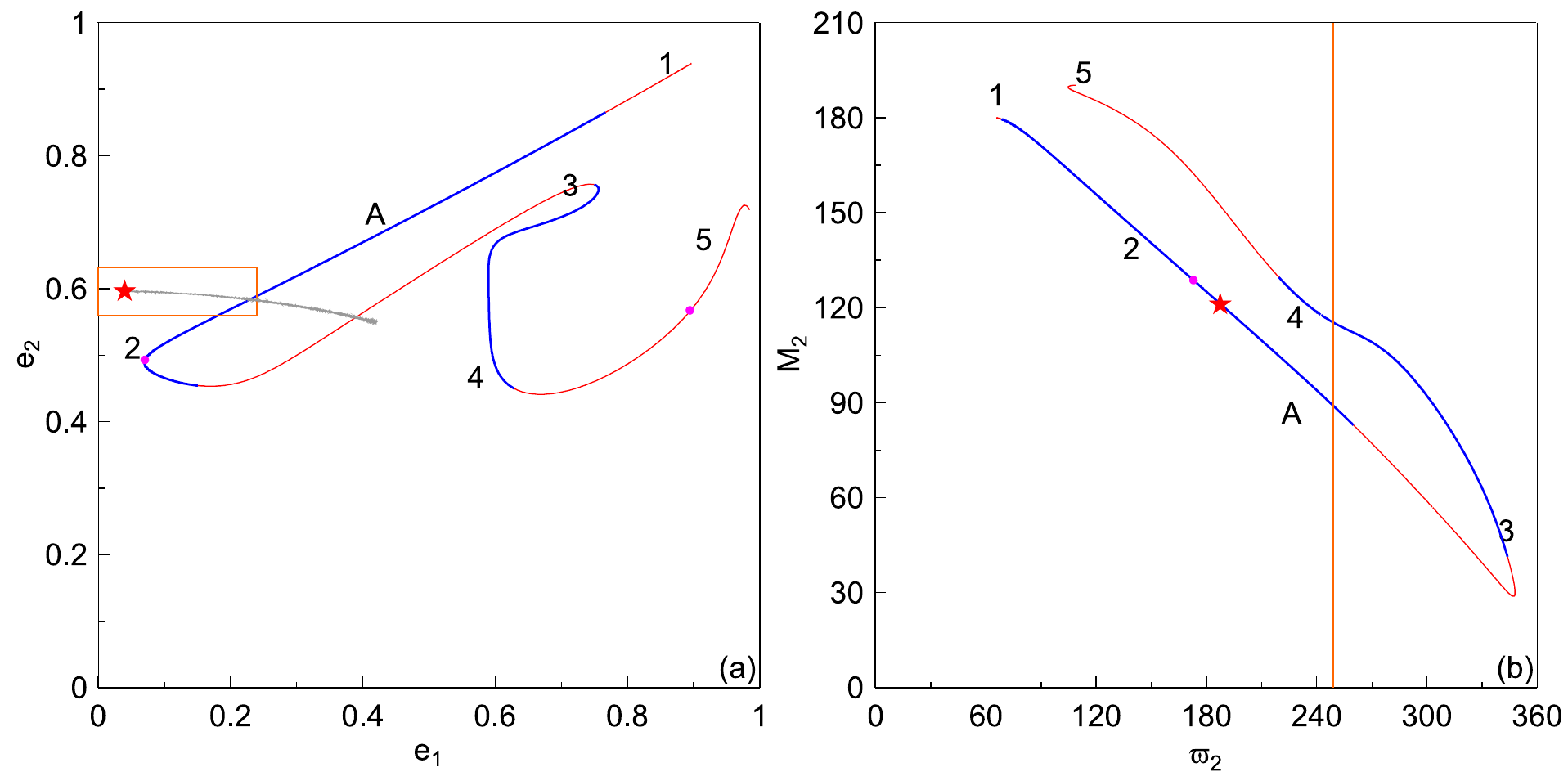}
\caption{\textbf{a} The asymmetric family $A$ of periodic orbits on $(e_1,e_2)$ plane computed for mass ratio $\rho\approx 2.22$ (HD 3651). The evolution of the system (red star indicates the initial position) is also likely to be evolving (gray curve) about an asymmetric periodic orbit. \textbf{b} The variation of $(\varpi_2,M_2)$ along the family $A$. Families are presented as in Fig. \ref{21}. Numbers are given to show the orientation of family $A$ in both panels.}
\label{3651as}
\end{figure}

Close to the dynamical location of HD 3651 and apart from the families of symmetric periodic orbits depicted in Fig. \ref{3651all}, we found a family, $A$, of asymmetric periodic orbits, which possesses a stable segment (see Fig. \ref{3651as}). A sample of this family is given in Table \ref{3651poa}. By using the suggested apsidal difference value ($187.5^\circ$) of Table \ref{tab}, we delve into family $A$ and seek for mean anomalies, which are not provided by the observations. By this means, we extract the value $M_2=121^\circ$ when $M_1=0^\circ$. By plugging these values into the rest of the suggested orbital elements and integrating the system, we get the evolution depicted by gray colour in Fig. \ref{3651as}a. It is centred at a stable asymmetric periodic orbit located at  $\alpha=1.58745$, $(e_1,e_2)=(0.227,0.582)$, $\Delta\varpi=97^\circ$ and $M_2=167^\circ$. 

\begin{table}
\caption{A representative sample of stable asymmetric periodic orbits of family $A$ related to the system HD 3651.}
\begin{center}
\begin{tabular}{ccccc}
\toprule
 $a_2/a_1$  &   $e_1$  &  $e_2$ &  $\Delta\varpi$ & $M_2$ \\
\midrule
 1.58738 &0.321 &0.629 &87.77 &172.18  \\ 
 1.58745 &0.227 &0.583 &97.11 &167.56  \\
 1.58754 &0.154 &0.546 &110.31 &160.80 \\
 1.58762 &0.112 &0.524 &124.63 &153.42 \\
 1.58768 &0.091 &0.512 &138.46 &146.32 \\
 1.58773 &0.078 &0.502 &152.82 &138.97 \\
 1.58776 &0.071 &0.495 &168.98 &130.70 \\
 1.58778 &0.069 &0.487 &186.40 &121.79 \\  
 1.58781 &0.073 &0.481 &201.25 &114.16 \\ 
 1.58787 &0.080 &0.475 &215.13 &107.00 \\
 1.58796 &0.092 &0.469 &229.26 &99.62  \\
 1.58809 &0.115 &0.461 &245.38 &91.01  \\
\bottomrule
\end{tabular}\end{center}
\label{3651poa}
\end{table}

\begin{figure}
\centering 
\includegraphics[width=\columnwidth]{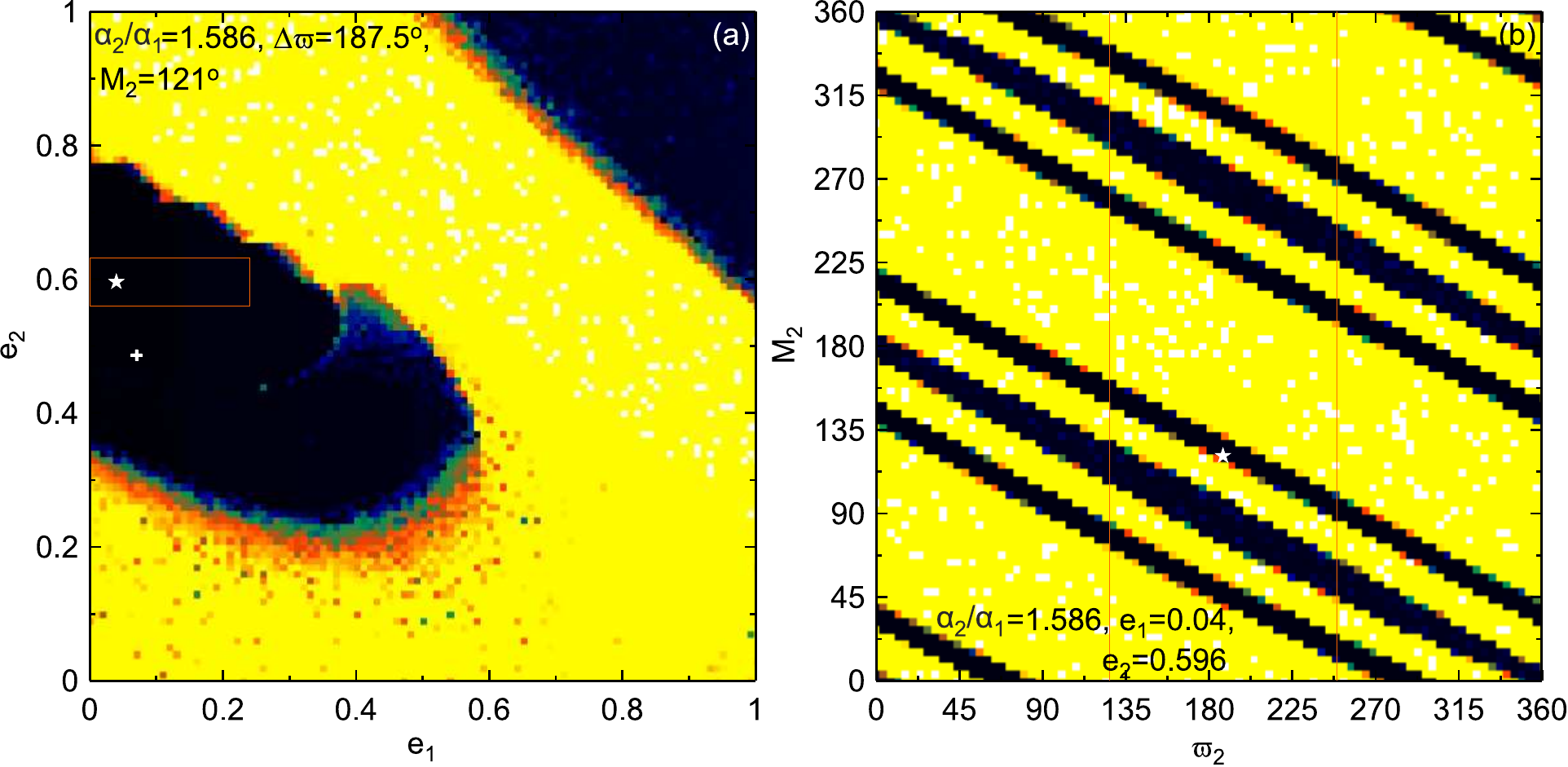} 
\caption{DS-map around a stable asymmetric periodic orbit (white cross) on the grid \textbf{a} $(e_1,e_2)$ and \textbf{b} $(\varpi_2,M_2)$. HD 3651 is depicted by the white star. The initial conditions that are fixed for the computation of the maps are shown on each panel.}
\label{3651awm}
\end{figure}

In Fig. \ref{3651awm}, we present DS-maps on grids $(e_1,e_2)$ and $(\Delta\varpi,M_2)$. In panel (a), we used as initial conditions the central values of the suggested semimajor axes and apsidal difference (see Table \ref{tab}) and derived the mean anomalies from the stable asymmetric periodic orbit (located at $(e_1,e_2)=(0.070,0.487)$, $\Delta\varpi=187.5^\circ$ and $M_2=121^\circ$) of the family $A$ presented in Fig. \ref{3651as}. We observe that this set creates an island of stability, which includes the possible range of the observational values indicated by the orange box. The observational eccentricities are depicted by the white star, while those of the periodic orbit by the white cross. We note that the boundaries of this island of stability did not change when we computed the DS-map for the semimajor axes taken from the above periodic orbit, i.e. when $\alpha=1.58779$. In panel (b), we chose the suggested semimajor axes and eccentricities and varied the longitude of pericentre and the mean anomaly of the outer planet. We observe that the regions of stability on this plane are narrow strips, which could justify the difficulty in tracing a fitting with two resonant planets evolving in the asymmetric configuration. These regions of stability restrict rigorously the deviations in longitudes of pericentre in relation to mean anomalies. 
 
Finally, along the previously discussed stable segment of family $A$, the vertical stability changes and a bifurcation point (vco) is created. The vco is located at $(e_1,e_2)=(0.070,0.493)$ and should generate spatial asymmetric periodic orbits, which under possible differential planetary migration may pump up the planetary inclinations.  

\subsection{HD 7449}

The suggested data for HD 7449 are shown in Table \ref{tab} \citep[see also][]{Dumu11}. The period ratio is $\frac{T_2}{T_1}\approx 3.17$. The masses are normalized as $m_1=0.001057$ and $m_2=0.0019047$. 

\begin{figure}
\centering
\includegraphics[width=\columnwidth]{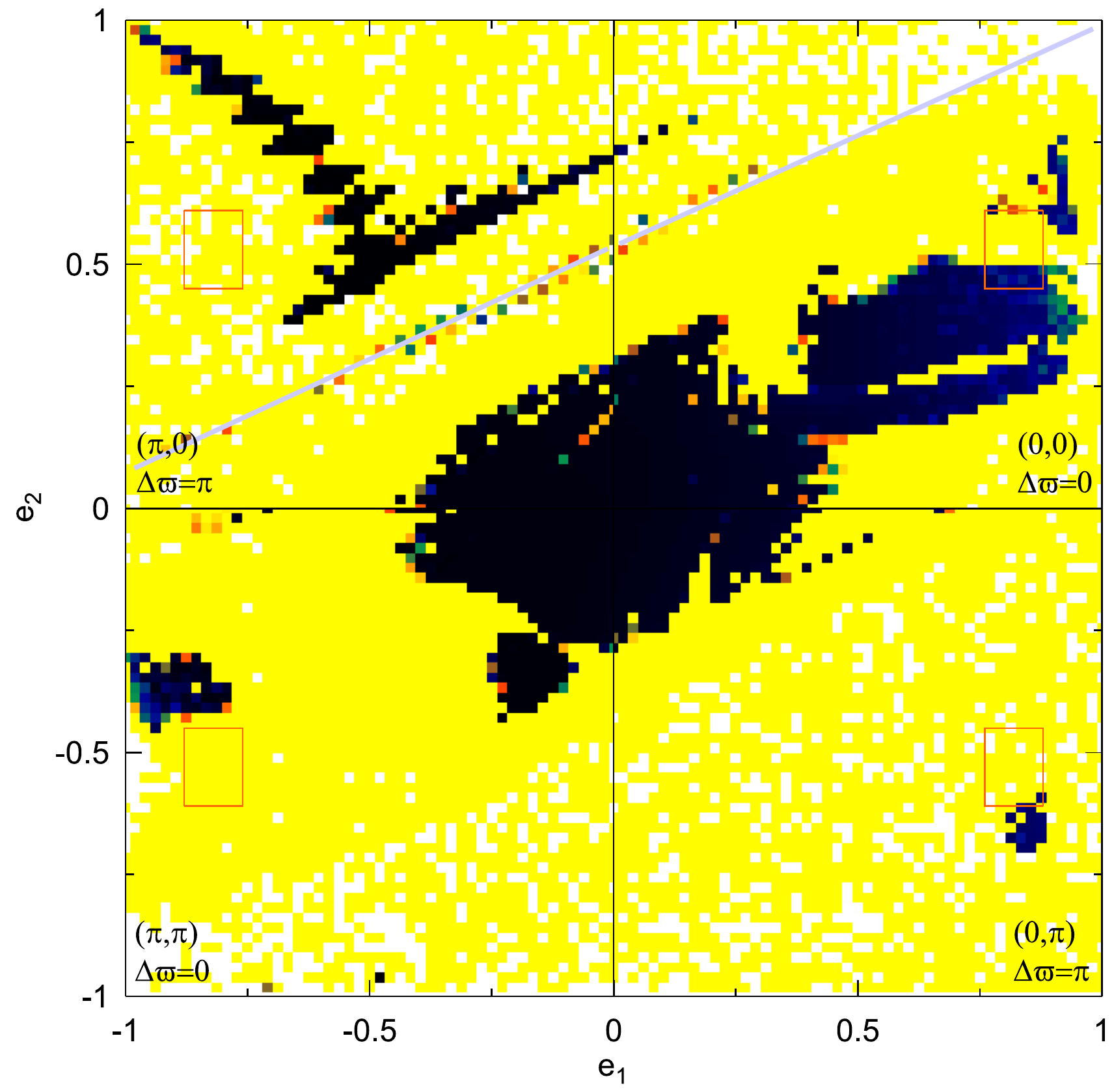}
\caption{DS-map on ($e_1,e_2$) plane of symmetric configurations computed for  $\alpha=2.156522$ and for the masses of HD 7449. Orange boxes indicate the values of eccentricities suggested by the observations.}
\label{7449all}
\end{figure}

In Fig. \ref{7449all}, we present the DS-map of symmetric configurations for $\alpha=2.156522$ (suggested ratio) and we observe that the $(0,0)$ configuration may allow the long-term stability for the suggested possible values of eccentricities.

\begin{figure}
\centering
\includegraphics[width=\columnwidth]{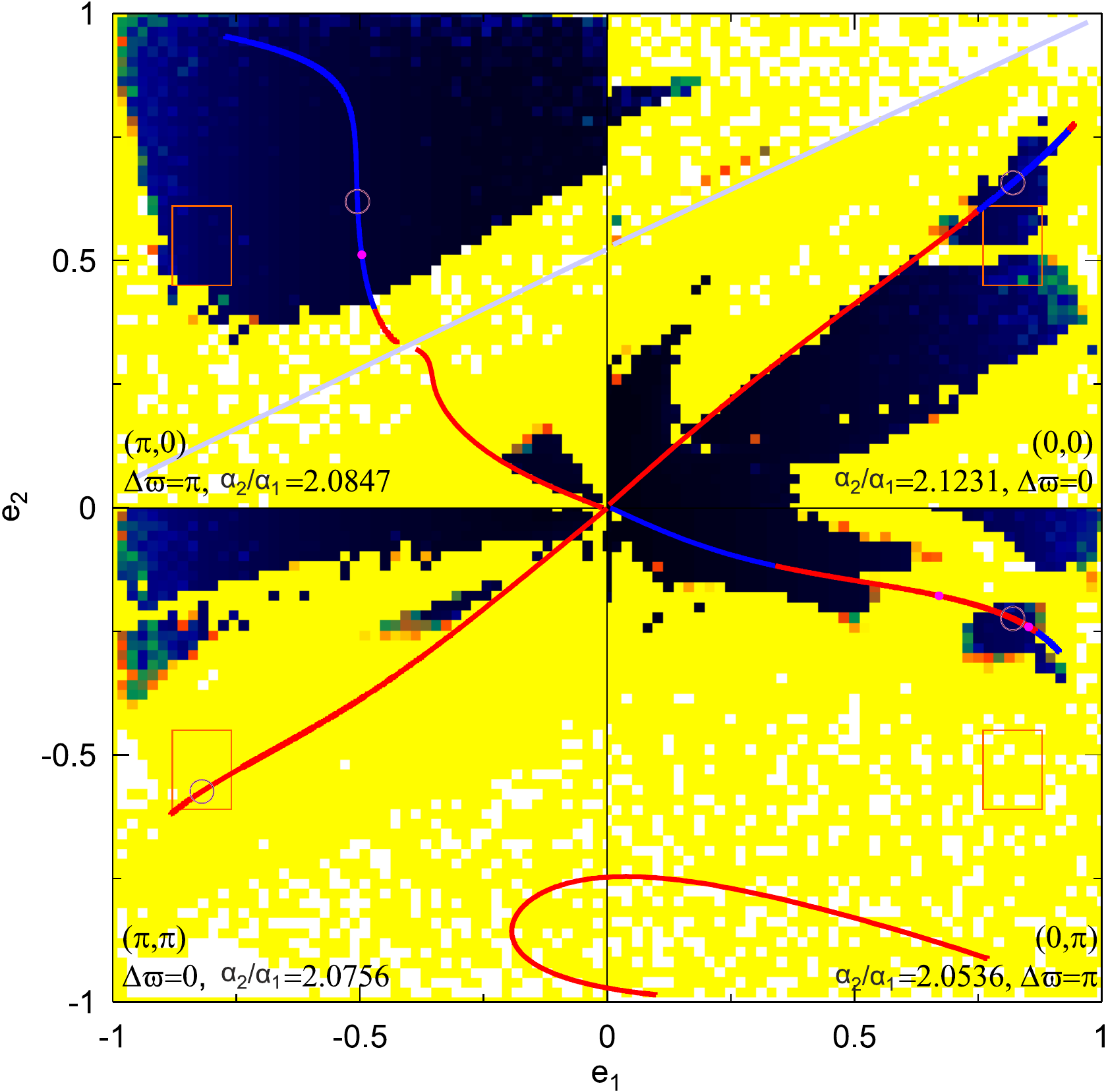}
\caption{DS-map on ($e_1,e_2$) plane of symmetric configurations computed for the indicated values $\alpha=a_2/a_1$, for each quadrant, which correspond to the exact resonances provided by periodic orbits (pink circles) and are close to the observational values ($e_1,e_2$) of the system HD 7449.}
\label{7449map}
\end{figure}

In Fig. \ref{7449map}, we provide another DS-map for each symmetric configuration, but we have chosen  initial values for the semimajor axis ratio $\alpha$ that correspond to exact resonances (periodic orbits)  at the  eccentricities $(e_1,e_2)=(0.820,0.657)$, $(e_1,e_2)=(-0.505,0.619)$, $(e_1,e_2)=(-0.819,-0.574)$ and $(e_1,e_2)=(0.820,-0.224)$ for each quadrant, respectively, depicted by pink circles. Stability regions related to the suggested eccentricities are now revealed for the $(\theta_3,\theta_1)$=($0,0$) and ($\pi,0$) configurations.  

\begin{figure}\centering
\includegraphics[width=\columnwidth]{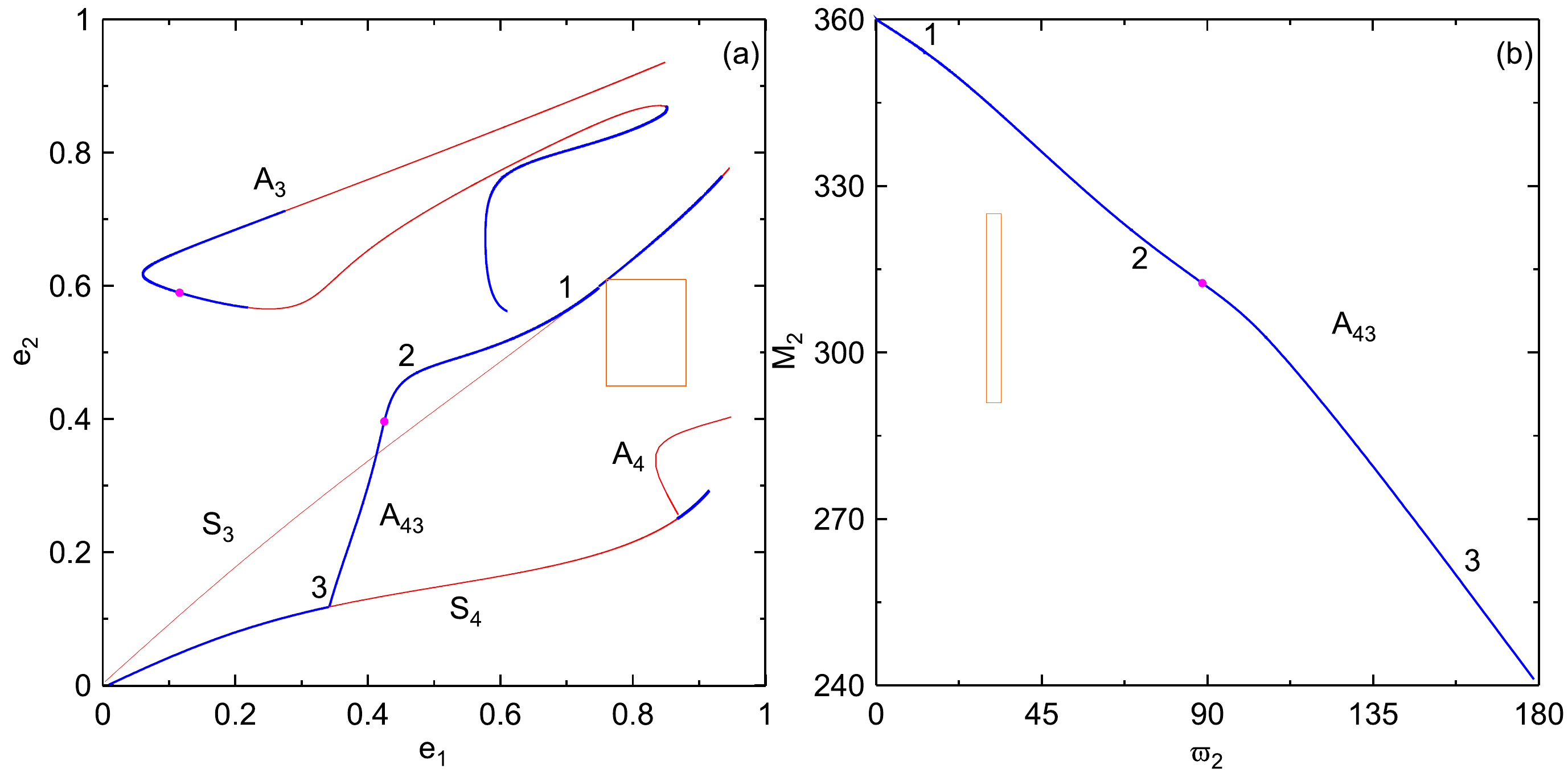}
\caption{\textbf{a} Families of symmetric ($S$) and asymmetric ($A$) periodic orbits on $(e_1,e_2)$ plane, which are related to the system HD 7449. \textbf{b} The family $A_{43}$ on the $(\varpi_2,M_2)$ plane. Orange box indicates the suggested position of HD 7449 according to Table \ref{tab} (presentation as in Fig. \ref{3651as}).}
\label{7449as}
\end{figure}

Nevertheless, in the neighbourhood (orange boxes) of these exoplanets as viewed on the eccentricities plane, we found a family with a segment of asymmetric stable periodic orbits, denoted as $A_{43}$, which bifurcates from the family of symmetric periodic orbits $S_4$ in the configuration $(0,\pi)$ and terminates at the bifurcation point of family $S_3$ in the configuration $(0,0)$ (see Fig. \ref{7449as}a). Also we computed the families $A_3$ and $A_4$. Many numerical computations performed have showed that family $A_3$ is not associated with regions of stability close enough to the observed planets and family $A_4$ is totally unstable. In Fig. \ref{7449as}b, we present family $A_{43}$ on the $(\varpi_2,M_2)$ plane. In Table \ref{7449poa}, we provide a sample of asymmetric periodic orbits of this family.

\begin{table}
\caption{A representative sample of the stable asymmetric periodic orbits (family $A_{43}$) in the dynamical vicinity of HD 7449 in phase space.}\begin{center}
\begin{tabular}{ccccc}
\toprule
 $a_2/a_1$  &   $e_1$  &  $e_2$ &  $\Delta\varpi$ & $M_2$ \\
\midrule
 2.09211 &0.350 &0.149 &138.08 &276.82   \\ 
 2.09449 &0.394 &0.277 &105.84 &302.76   \\
 2.09578 &0.422 &0.382 &93.04 &310.33   \\
 2.09864 &0.475 &0.471 &69.29 &322.11   \\
 2.10084 &0.530 &0.491 &56.33 &329.39   \\
 2.10451 &0.614 &0.521 &39.37 &339.58   \\
 2.10737 &0.669 &0.548 &27.92 &346.37   \\
 2.10959 &0.709 &0.571 &18.78 &351.39    \\  
 2.11117 &0.736 &0.589 &10.09 &355.63    \\ 
 2.11185 &0.748 &0.599 &0.36 &359.85    \\
\bottomrule
\end{tabular}\end{center}
\label{7449poa}
\end{table}

\begin{figure}
\centering 
\includegraphics[width=\columnwidth]{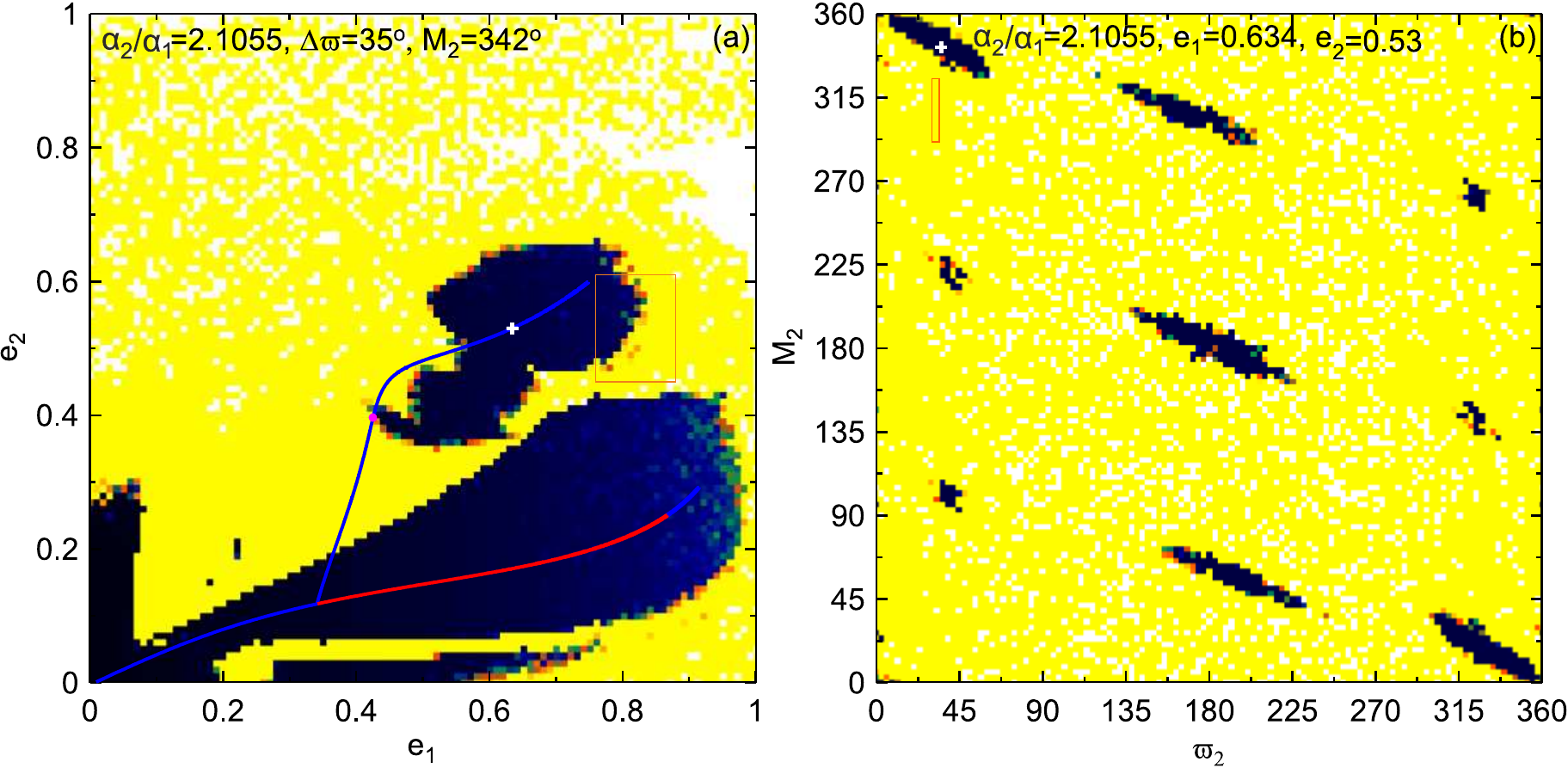}
\caption{DS-maps computed on \textbf{a} $(e_1,e_2)$ and \textbf{b} $(\varpi_2,M_2)$ grids for HD 7449. The initial conditions that are fixed for the computation of the maps are shown on each panel. Orange boxes represent the possible position of the system according to Table \ref{tab}.}
\label{7449awm}
\end{figure}

Two-planet fittings for this exosystem imposed a decrease in the eccentricities $e_i\leq0.2$, in order for long-term stability to be achieved \citep{wit13}. Given the family of stable asymmetric periodic orbits, we can suggest an alternative  configuration that could guarantee the dynamical stability for very long timescales. Indeed, the DS-map of Fig. \ref{7449awm}a reveals the existence of an island of stability around an asymmetric periodic orbit. We constructed the map by using as initial conditions those of a stable asymmetric periodic orbit of the family $A_{43}$ which is located close to the observational data. It corresponds to $\alpha=2.105516$, $e_1=0.634$, $e_2=0.530$, $\varpi_2=35^{\circ}$ and $M_2=342^{\circ}$. The island of stability is located slightly in lower $e_1$ values than those given in Table \ref{tab} and, therefore a suggestion for a lower value of $e_1$ for the system is reasonable. In Fig. \ref{7449awm}b, we construct a DS-map on the plane ($\varpi_2, M_2$) with the rest fixed conditions taken from the periodic orbit mentioned above. We observe that the possible regions of stability on the plane $(\varpi_2,M_2)$ are confined to islands. One of them seems to be related to the position of the HD 7449, but the DS-map suggests a reasonable correction for the real position of the system. 

\subsection{HD 89744}

The suggested data for HD 89744 are shown in Table \ref{tab} \citep[see also][]{sylvain00}. The period ratio is $\frac{T_2}{T_1}\approx 3.005$. The normalized planetary masses are $m_1=0.002268$ and $m_2=0.005104$. 

\begin{figure}
\centering
\includegraphics[width=\columnwidth]{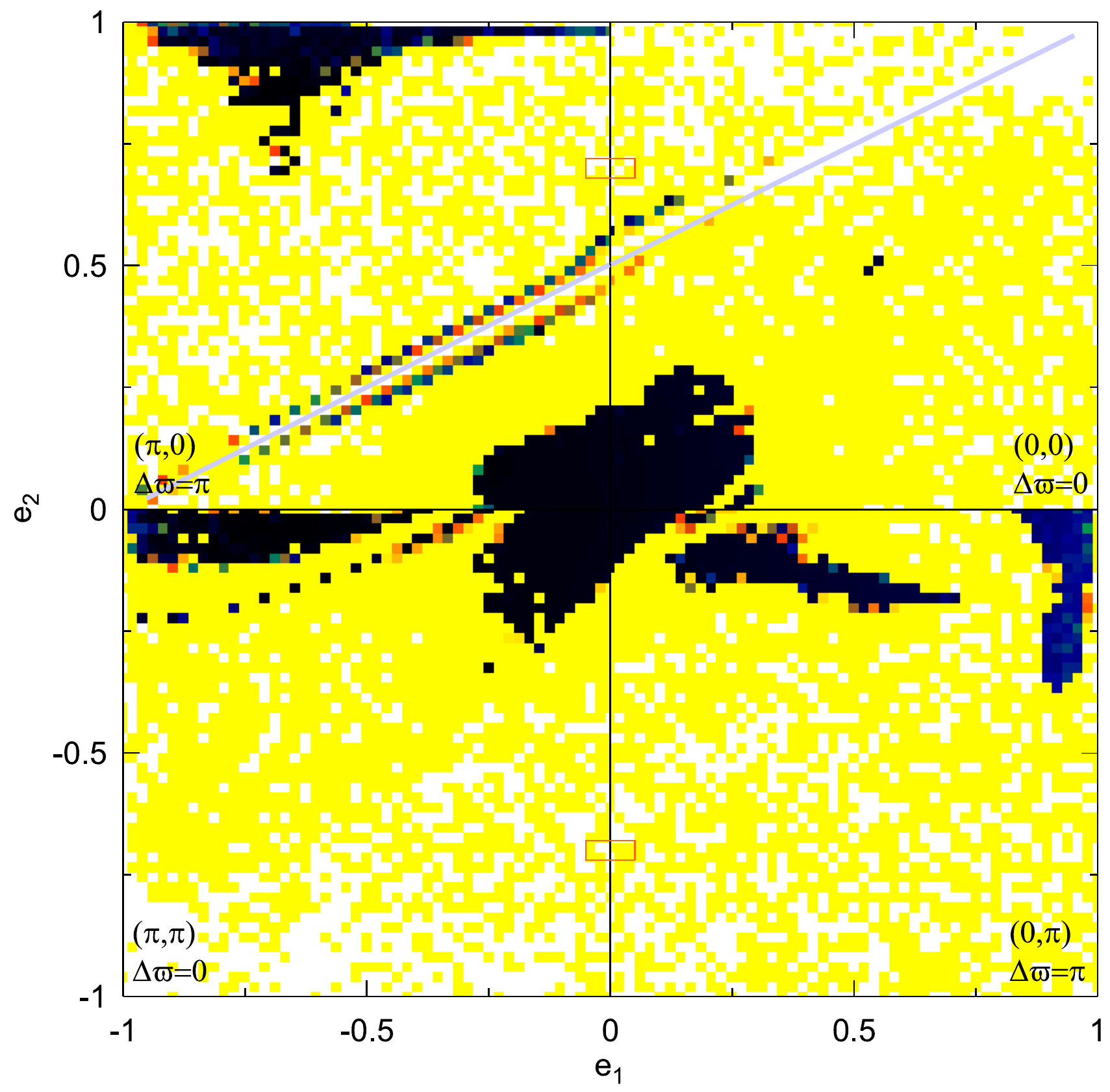}
\caption{DS-map on ($e_1,e_2$) plane of symmetric configurations computed for  $\alpha=2.0$ and for the planetary masses of system HD 89744. Orange boxes indicate the values of eccentricities suggested by the observations.}
\label{89744all}
\end{figure}

\begin{figure}
\centering
\includegraphics[width=\columnwidth]{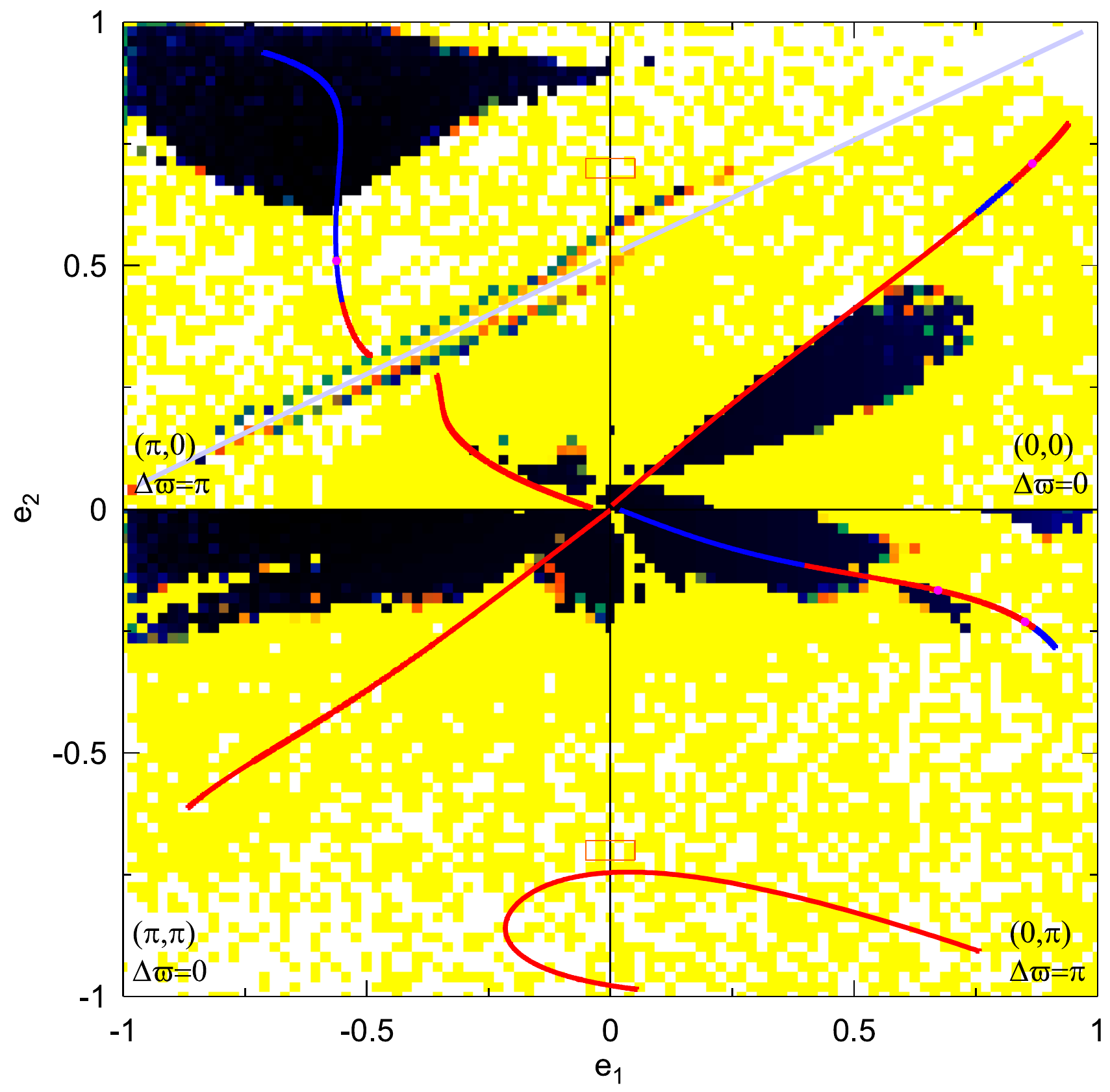}
\caption{DS-map as in Fig. \ref{89744all}, but for $\alpha=3^{2/3}\approx 2.08$. The $3$:$1$ resonant families of symmetric periodic are also presented.}
\label{89744map}
\end{figure}

In the DS-map of Fig. \ref{89744all}, which has been computed for the symmetric configurations for $\alpha=2.0$ (see Table \ref{tab}), we observe that the suggested high eccentricity of the outer planet locates the system in an unstable dynamical region. If we recompute the map for the Keplerian position of the resonance, $\alpha=3^{2/3}\approx 2.08$, we get the DS-map of Fig. \ref{89744map}, where also the system is located in chaos. We note that close to MMR, slight variations of the semimajor axis ratio may change significantly the distribution of regular and chaotic regions.  

\begin{figure*}\centering
\includegraphics[width=2.1\columnwidth]{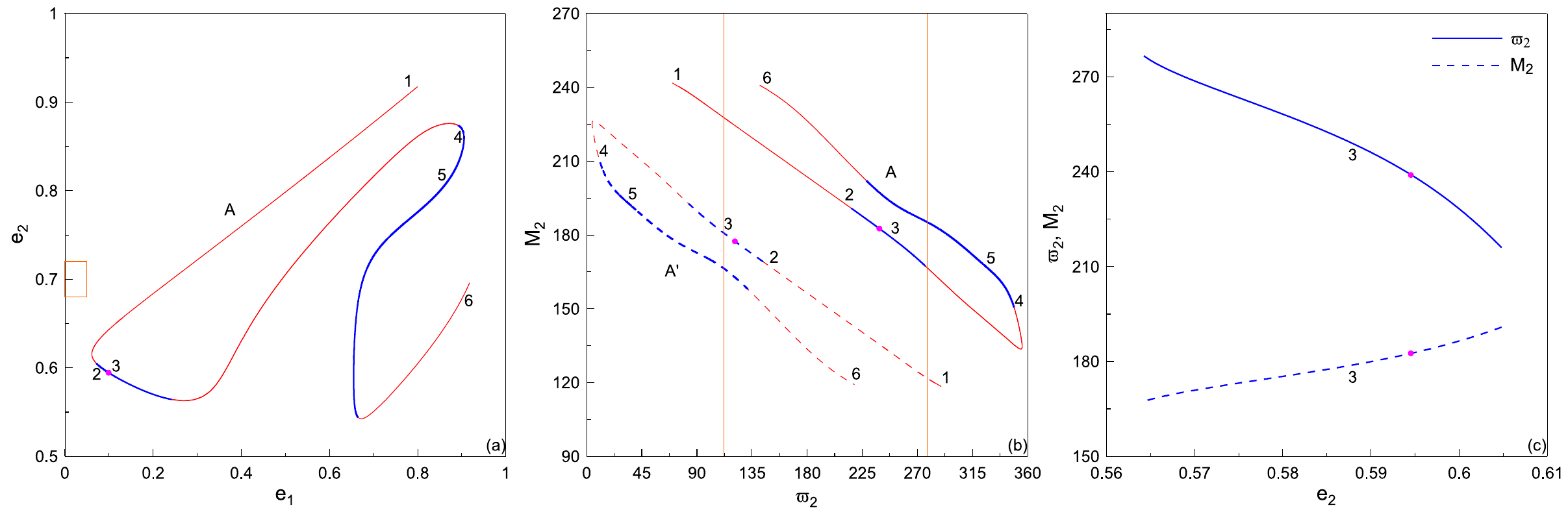}
\caption{The family $A$ of periodic orbits related to the system HD 89744 shown on \textbf{a} $(e_1,e_2)$ plane and \textbf{b} $(\varpi_2,M_2)$ plane, where the mirror image of the family $A'$ is also presented. \textbf{c} Variation of $\varpi_2$ and $M_2$ along the stable segment of family $A$ (indicated by the domain 3) using $e_2$ as a parameter.}
\label{89744as}
\end{figure*}

Since symmetric configurations were thus excluded, we sought for asymmetric periodic orbits in the dynamical vicinity of the system around the suggested data. Indeed, we found a family ($A$) of them, which is presented in Fig. \ref{89744as} and a sample of it is given in Table \ref{89744poa}. 

\begin{table}
\caption{A representative sample of the stable asymmetric periodic orbits of family $A$ in the dynamical vicinity of HD 89744 in phase space.}
\begin{center}
\begin{tabular}{ccccc}
\toprule
 $a_2/a_1$  &   $e_1$  &  $e_2$ &  $\Delta\varpi$ & $M_2$ \\
\midrule
 2.09011 &0.072 &0.605 &216.12 &190.81  \\ 
 2.08975 &0.077 &0.602 &222.26 &188.62  \\
 2.08932 &0.084 &0.599 &228.74 &186.30  \\
 2.08885 &0.092 &0.596 &235.15 &183.98  \\
 2.08834 &0.104 &0.593 &241.46 &181.67  \\
 2.08781 &0.118 &0.589 &247.63 &179.37  \\
 2.08726 &0.136 &0.584 &253.59 &177.10  \\
 2.08671 &0.157 &0.579 &259.25 &174.85  \\  
 2.08616 &0.180 &0.574 &264.58 &172.66  \\ 
 2.08562 &0.205 &0.569 &269.55 &170.53  \\
\bottomrule
\end{tabular}\end{center}
\label{89744poa}
\end{table}

\begin{figure}\centering 
\includegraphics[width=\columnwidth]{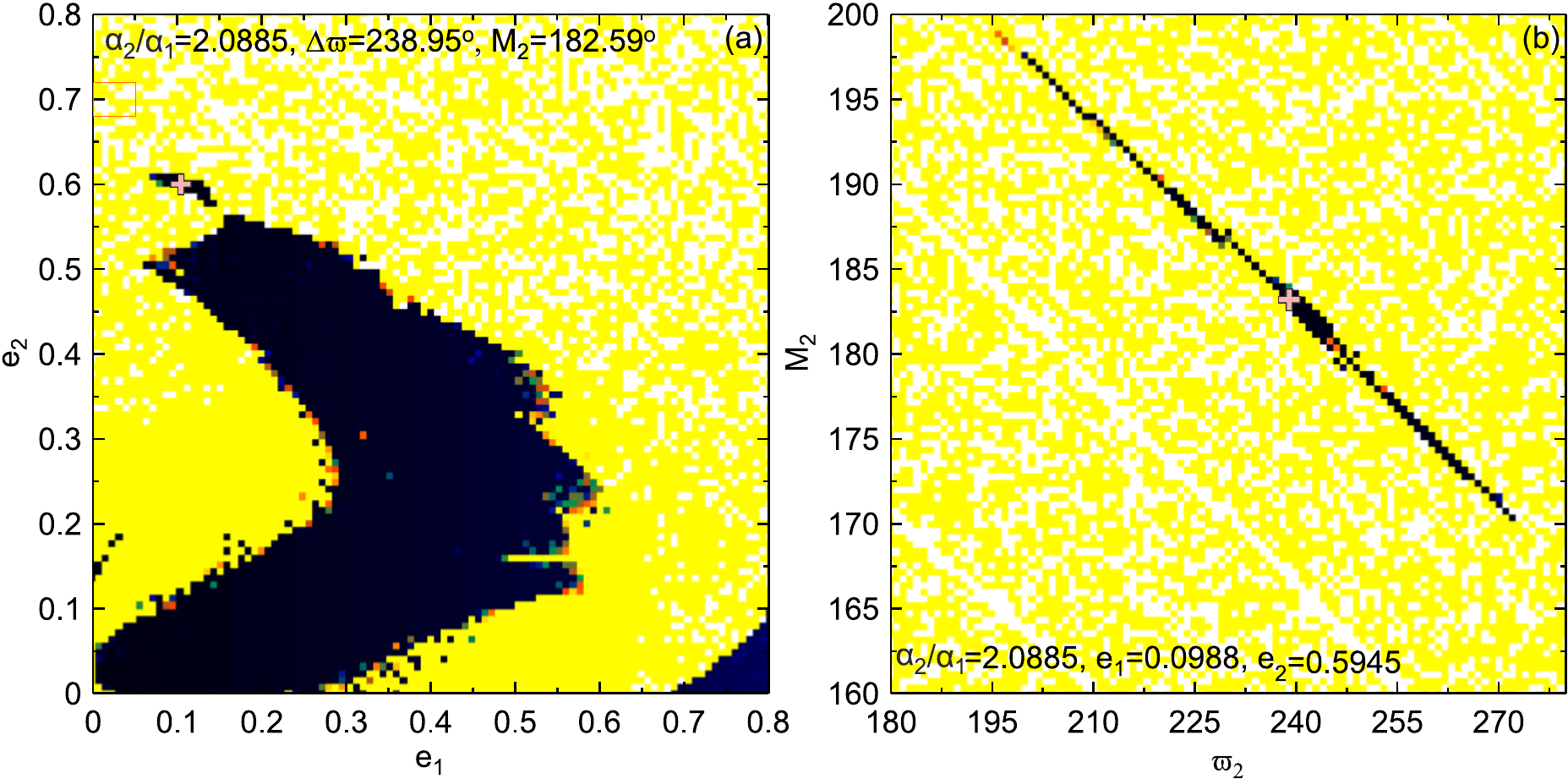}
\caption{DS-maps for an asymmetric configuration of the system HD 89744 based on the stable ACR depicted by the pink cross on the grids \textbf{a} $(e_1,e_2)$ and \textbf{b} $(\varpi_2,M_2)$. The elements that remain fixed for the computation of the grids are shown on each panel.}
\label{89744awm}
\end{figure}

In order to determine stable regions for large $e_2$ values, we consider the initial conditions of an asymmetric orbit at ($e_1,e_2$)=($0.099,0.595$) with $\alpha=2.08855$, $\varpi_2=238.95^{\circ}$, and $M_2=182.59^{\circ}$ (a reference ACR). By varying the eccentricities and keeping fixed the rest initial conditions we obtain the DS-map of Fig. \ref{89744awm}a. There exists a large region of stable orbits, which extends to about $e_2\approx 0.55$. By choosing different reference ACRs we could not extend this region to larger $e_2$ values, i.e. up to the corresponding observational value of HD 89744b. Also, by computing the DS-map by varying $\varpi_2$ and $M_2$, we obtain that the stability region appears to be restricted in a very thin zone (Fig. \ref{89744awm}b). Thus, when high eccentricity values of $e_2$ are considered, the stability region around the asymmetric ACR is very sensitive to the variation of the planetary phases. We mention that in two-planet fittings of \citet{wit13} a significant decrease in the eccentricities ($e_i\approx 0.2$) was imposed, in order for long-term stability to be achieved.

\subsection{HD 102272}

The suggested data for HD 102272 are shown in Table \ref{tab} \citep[see also][]{nied09}. The period ratio is $\frac{T_2}{T_1}\approx 4.075$. The planetary masses are normalized as $m_1=0.003091$ and $m_2=0.001362$. 

\begin{figure}
\centering
\includegraphics[width=\columnwidth]{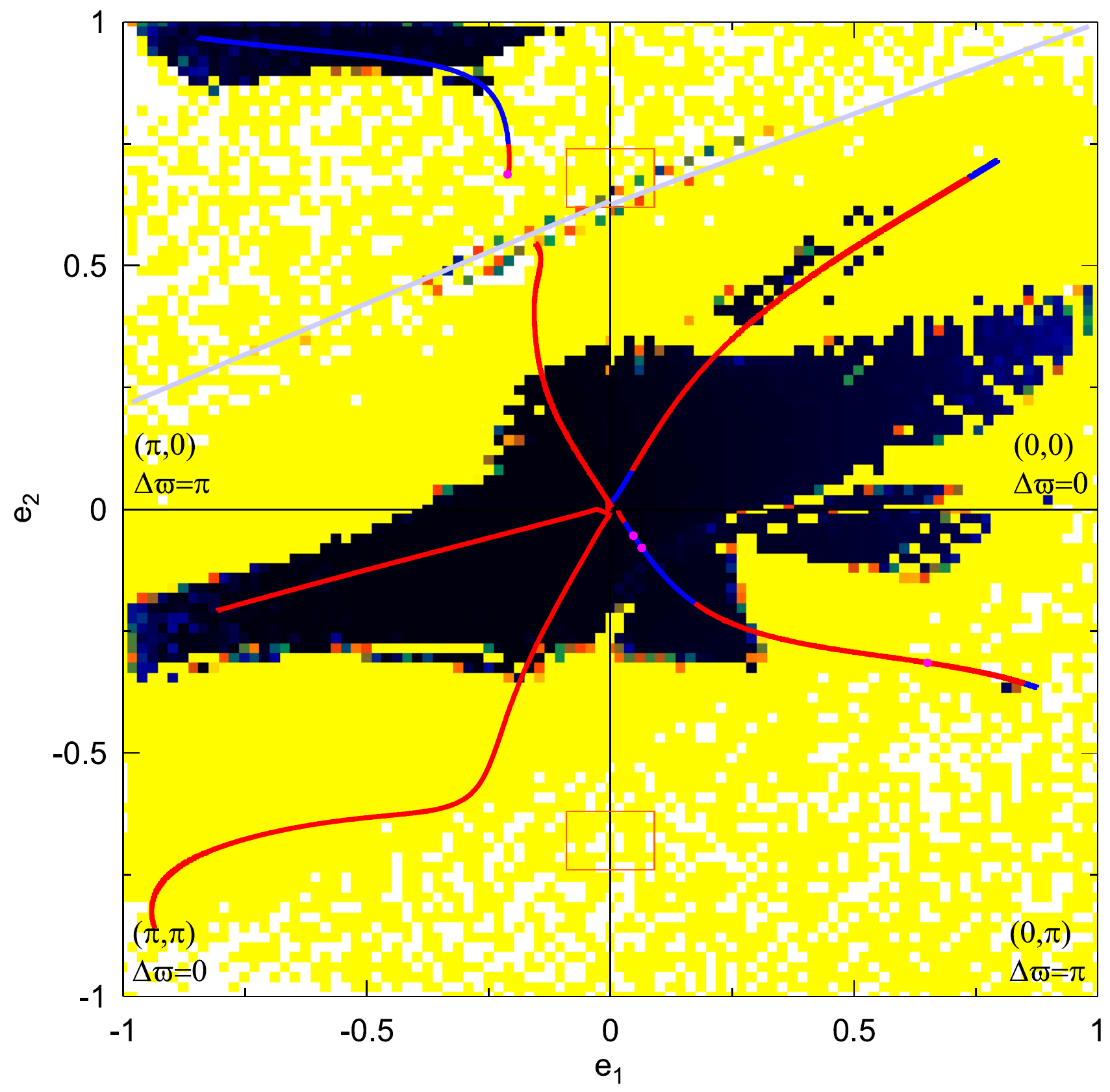}
\caption{DS-map of symmetric configurations for $\alpha=2.557$ and for the planetary masses of system HD 102272. The families of $4$:$1$ resonant families of periodic orbits are also presented. Orange boxes indicate the values of eccentricities suggested by the observations.}
\label{102272map}
\end{figure}

\begin{figure*}\centering
\includegraphics[width=2.1\columnwidth]{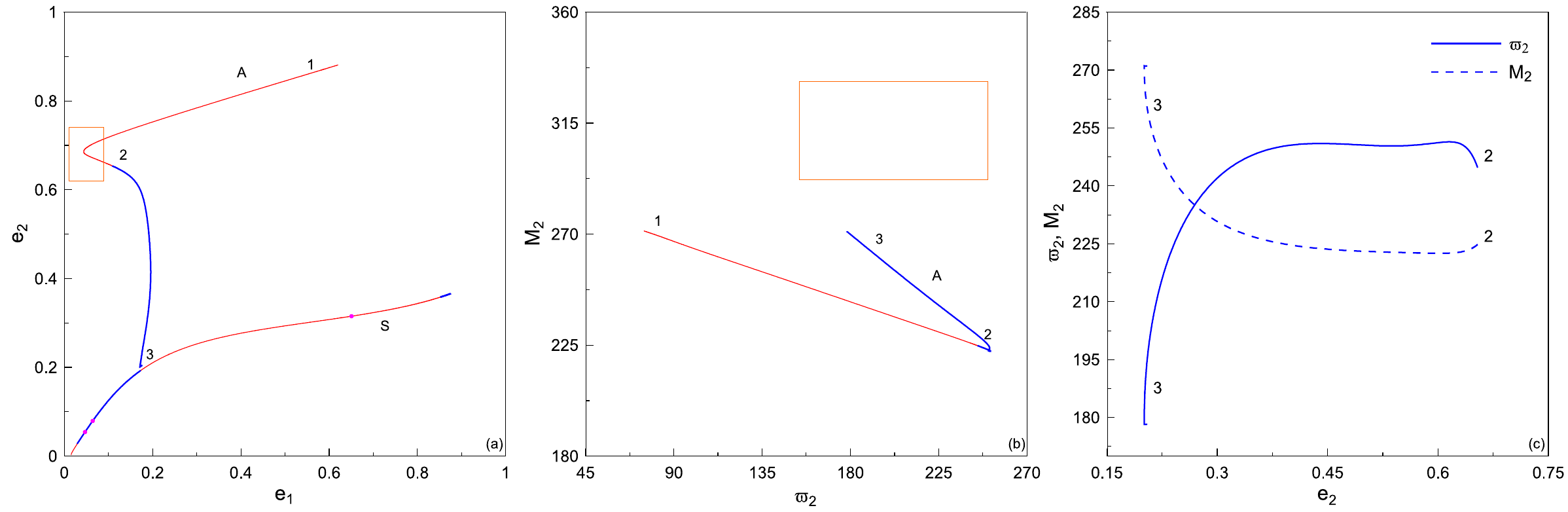}
\caption{The asymmetric family $A$ related to the system HD 102272 \textbf{a} projection in the $(e_1,e_2)$ plane. The symmetric family $S$, from which family $A$ bifurcates, is also presented.  \textbf{b} Projection on the $(\varpi_2,M_2)$ plane and \textbf{c} variation of $\varpi_2$ and $M_2$ along the family using $e_2$ as a parameter. The orange box indicates the domain of initial conditions given in Table \ref{tab}.}
\label{102272as}
\end{figure*}

The DS-map presented in Fig. \ref{102272map} was computed for the observational semimajor axis ratio $\alpha=2.557$ definitely precludes long-term stability, in case these planets are evolving in a symmetric configuration. However, we have computed a family of asymmetric periodic orbits, $A$, with a stable segment in the dynamical vicinity of these planets. The family is presented, in different planes, in Fig. \ref{102272as} and a sample of it is given in Table \ref{102272poa}.

\begin{table}
\caption{A representative sample of the stable asymmetric periodic orbits in the dynamical vicinity of HD 102272 in phase space.}\begin{center}
\begin{tabular}{ccccc}
\toprule
 $a_2/a_1$  &   $e_1$  &  $e_2$ &  $\Delta\varpi$ & $M_2$ \\
\midrule
 2.52079 &0.108 &0.654 &244.75 &224.82 \\ 
 2.51987 &0.128 &0.644 &248.55 &223.67 \\
 2.51953 &0.192 &0.499 &250.57 &223.05 \\
 2.52116 &0.189 &0.316 &244.62 &229.13 \\
 2.52348 &0.182 &0.269 &235.21 &234.95 \\
 2.52601 &0.177 &0.242 &225.19 &241.07 \\
 2.52868 &0.175 &0.223 &214.68 &247.60  \\
 2.53170 &0.173 &0.210 &202.95 &255.05 \\  
 2.53484 &0.172 &0.203 &190.99 &262.77 \\ 
 2.53832 &0.172 &0.201 &178.22 &271.08 \\
\bottomrule
\end{tabular}\end{center}
\label{102272poa}
\end{table}

\begin{figure}
\centering 
\includegraphics[width=\columnwidth]{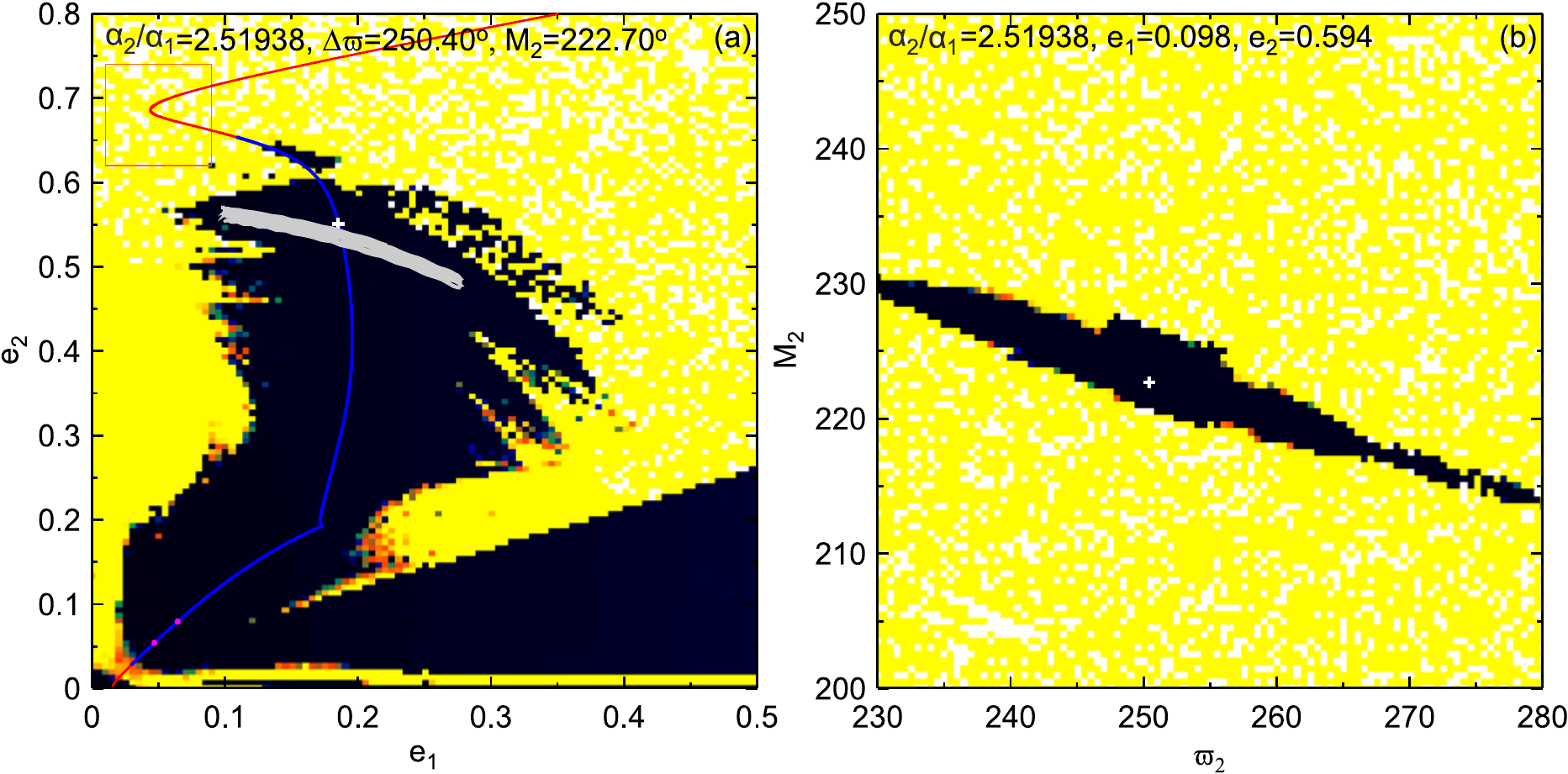}
\caption{DS-maps for an asymmetric configuration of the system HD 102272 based on the stable ACR depicted by the white cross on the grids \textbf{a} $(e_1,e_2)$ and \textbf{b} $(\varpi_2,M_2)$. The elements that remain fixed for the computation of the grids are shown on each panel.}
\label{102272awm}
\end{figure}

Similarly to the previous cases, we consider a stable periodic orbit along the family with eccentricities close to those of Table \ref{tab}. Particularly, we choose as a reference ACR the periodic orbit of family $A$ at $e_1=0.185$, $e_2=0.551$ with $\alpha=2.519388$, $\varpi_2=250.40^{\circ}$ and $M_2=222.70^{\circ}$. We obtain from Fig. \ref{102272as}c that the reference periodic orbit is located in a plateau for the values $\varpi_2$ and $M_2$ with respect to the eccentricity $e_2$, while $e_1$ does not vary significantly along this segment. The DS-map of Fig. \ref{102272awm}a reveals the existence of a large stability region. Orbits starting in this domain show oscillations about asymmetric periodic orbits of the family $A$. An example of such an evolution projected on the plane of eccentricities is shown with the gray line. The eccentricity amplitude variations are reduced as the initial conditions approach an orbit of the family $A$ (i.e. the asymmetric ACR). However, it is clearly seen that the initial eccentricities of HD 102272 suggested by the fitting of the observational data (orange box) is located above the stability region and inside the sea of strong chaos. The DS-map of Fig. \ref{102272awm}b restricts the stability in particular values of planetary phases around those defined by the reference periodic orbit.

\section{Conclusions}
In the present work, we study the dynamical stability of two-planet systems, modelled by the general three body problem, which possess at least one planet evolving on a highly eccentric orbit. The criterion for long-term stability is the regular evolution of the orbits, which is concluded after computing chaotic indices and particularly, the de-trended Fast Lyapunov Indicator (DFLI). Regular evolution with high eccentricities can be established inside apsidal and mean motion resonances (MMR). The exact position of MMR at moderate or high eccentricities is given by families of elliptic periodic orbits or, equivalently, apsidal corotation resonances (ACR).  
 
By taking into account the collision criterion of orbits (in the unperturbed approximation) we can separate the space of initial conditions in different domains, $D_i$ ($i=0,..,3$). Regular orbits can be found in all such domains, as it is revealed by our computation of maps of dynamical stability for a fictitious Jupiter-Saturn system. In the domain $D_0$, where unperturbed orbits cannot cross, the majority of initial conditions, for which the inner planet starts with a moderate eccentricity ($e_1=0.3$) and is aligned with the outer one ($\Delta\varpi(0)=0$), shows regular evolution, when $\alpha>1.7$ (i.e. to the right of the 2$:$1 MMR). By increasing the eccentricity of the inner planet ($e_1=0.6$) a large part of the above mentioned regular region extends to the domain $D_1$, where anti-aligned orbits collide. MMRs and apsidal resonances ($\Delta\varpi$ librations around zero) provide the necessary conditions for long-term stability. The above dynamical picture is strongly affected when we start the planets  with anti-aligned apsides ($\Delta\varpi(0)=\pi$). The area of stability region decreases rapidly and gets confined mainly to islands, each of which is associated with a particular MMR.  

When planetary orbits are so elongated that cross each other in spite of their apsidal angle (domain $D_2$), then, the necessary condition for long-term stability is provided only by MMR. Now the regular regions are confined to islands of resonant motion which become evident either as apsidal corotation resonances (ACR) or as resonances between the average frequencies of rotation and libration of the resonant angles $\theta_1$ and $\theta_2$. 

The exosolar planets HD 82943b,c consist an indicative example of evolution inside the $(0,0)$ symmetric configuration of the $2$:$1$ MMR. The DS-maps show the existence of stability regions which extend along and around the corresponding family of periodic orbits. Although the masses of the planets are relatively large, the stability domain is quite wide for hosting the system and ensuring its long-term stability.  

In the system HD 3651, the inner planet evolves on an almost circular orbit. Besides the proposed data fittings that indicate low eccentricities for both planets \citep{wit13}, the single-planet fitting \citep{wit09} indicates that the outer planet may be very eccentric. Our dynamical analysis proves that this is possible. We showed the existence of a large region of stability for the $(\pi,0)$ symmetric configuration, but the system is located close to the border of this region and hence, it evolves with large amplitude eccentricity oscillations. Additionally, an island of stability, which potentially includes the position of the system, has been found around an asymmetric ACR. The systems HD 89744 and HD 102272 show similar features as those of HD 3651. For each case, we found an asymmetric ACR that neighbours the suggested position of the system. However, the corresponding stability areas around the ACR are located slightly below the eccentricity value $e_2$ of the outer planet given in Table \ref{tab}. Also, the stability areas are restricted to very narrow strips on the plane of phases $(\Delta\varpi,M_2)$. Therefore, the existence of systems in such an asymmetric configuration is possible, but maybe not so probable. 

In the system HD 7449 the orbits of both planets are very eccentric. Lower eccentricities are proposed in \citep{wit13} in order for stability to be ensured. In our dynamical approach, by considering that the system is locked in the $3$:$1$ MMR, we determined three possible stability domains at high eccentricities that may host the system. One is associated with the symmetric $(\pi,0)$ configuration, however, the system is located close to the border of the region and far from the corresponding ACR. With a small increase of the eccentricity of the outer planet we can locate the system close to $(0,0)$ symmetric ACR.  The existence of a family of asymmetric periodic orbits (family $A_{43}$) guided us to detect another significant stability region. In this case, the phases ($M_2$ and $\Delta\varpi$) should be restricted in a well defined island in the corresponding DS-map.

The study of dynamical stability restricts significantly the possible configurations of planetary systems with high eccentric orbits. Nevertheless, long-term stability of such systems is possible. The mean motion resonance is a necessary dynamical mechanism for the long-term stability. The exact position of mean motion resonances at high eccentricities can be determined by the computation of periodic orbits, which may be helpful for constraining the uncertainties coming from the fitting analysis of observational data. Generally, the detected stability regions, even if they are broad or narrow, they cannot assure the existence of a system in any position of these regions. Additionally, such potential positions should be derived as reliable solutions from the fittings of sufficient and valid observational data \citep[see e.g.][]{corr05}. In our study, we focused on searching for stable configurations (particularly, stable periodic orbits) close to solutions suggested by particular studies of observational data.

%\section*{Acknowledgements} 

\bibliographystyle{mnras}
\bibliography{nbib}

% Don't change these lines
\bsp	% typesetting comment
\label{lastpage}
\end{document}